\documentclass[acmsmall, table]{acmart}

\usepackage{xcolor}

\usepackage{booktabs}
\usepackage{multirow}
\usepackage{makecell}
\usepackage{graphicx}
\usepackage{amsthm}

\usepackage{subcaption}
\usepackage{lscape}
\usepackage{tikz}
\usepackage[edges]{forest}
\usetikzlibrary{trees, positioning, shapes, shadows, arrows.meta}
\usepackage{amsmath}
\usepackage{tabularray}
\usepackage{tabularx}
\usepackage{bbding}
\usepackage{array}
\usepackage{algorithm}
\usepackage{algorithmic}

\definecolor{f_beige}{RGB}{255,249,221}
\definecolor{f_orange}{RGB}{255,225,199}
\definecolor{f_green}{RGB}{230,234,196}
\definecolor{f_purple}{RGB}{222,207,227}
\definecolor{f_sapphire}{RGB}{114,188,213}
\definecolor{f_azure}{RGB}{082,143,173}
\definecolor{f_stone}{RGB}{055,103,149}
\definecolor{f_navyblue}{RGB}{030,070,110}
\definecolor{f_preretrieval}{RGB}{217,234,211}
\definecolor{f_retrieval}{RGB}{174,65,50}
\definecolor{f_postretrieval}{RGB}{173,111,195}
\definecolor{f_generation}{RGB}{96,169,225}

\AtBeginDocument{%
  }

\setcopyright{acmlicensed}
\copyrightyear{2018}
\acmYear{2018}
\acmDOI{XXXXXXX.XXXXXXX}

\acmConference[Conference acronym 'XX]{Make sure to enter the correct
  conference title from your rights confirmation emai}{June 03--05,
  2018}{Woodstock, NY}
\acmISBN{978-1-4503-XXXX-X/18/06}

\begin{document}

\title{The Survey of Retrieval-Augmented Text Generation in Large Language Models}

\author{Yizheng Huang}
\email{hyz@yorku.ca}
\author{Jimmy X. Huang}
\email{jhuang@yorku.ca}
\affiliation{%
  \institution{York University}
  \city{Toronto}
  \state{Ontario}
  \country{Canada}
}

\renewcommand{\shortauthors}{Huang et al.}

\begin{abstract}
Retrieval-Augmented Generation (RAG) merges retrieval methods with deep learning advancements to address the static limitations of large language models (LLMs) by enabling the dynamic integration of up-to-date external information. This methodology, focusing primarily on the text domain, provides a cost-effective solution to the generation of plausible but possibly incorrect responses by LLMs, thereby enhancing the accuracy and reliability of their outputs through the use of real-world data. As RAG grows in complexity and incorporates multiple concepts that can influence its performance, this paper organizes the RAG paradigm into four categories: pre-retrieval, retrieval, post-retrieval, and generation, offering a detailed perspective from the retrieval viewpoint. It outlines RAG's mechanics and discusses the field's progression through the analysis of significant studies. Additionally, the paper introduces evaluation methods for RAG, addressing the challenges faced and proposing future research directions. By offering an organized framework and categorization, the study aims to consolidate existing research on RAG, clarify its technological underpinnings, and highlight its potential to broaden the adaptability and applications of LLMs.
\end{abstract}



\received{20 February 2007}
\received[revised]{12 March 2009}
\received[accepted]{5 June 2009}

\maketitle

\section{Introduction}
\label{sec:intro}
The advent of ChatGPT has significantly impacted both academia and industry due to its interactive capabilities and widespread application, establishing itself as a leading artificial intelligence tool \cite{DBLP:journals/corr/abs-2305-18486, DBLP:journals/corr/abs-2306-04504, DBLP:journals/corr/abs-2402-11203}. At the core of ChatGPT is the large language model (LLM) GPT-4, as detailed by \cite{achiam2023gpt}, which has seen numerous enhancements to its predecessors, showcasing exceptional abilities in a variety of Natural Language Processing (NLP) tasks \cite{DBLP:conf/ai/LaskarHH20}. Despite these advancements, the adoption of LLMs has highlighted several critical issues primarily due to their reliance on extensive datasets. This reliance restricts their ability to incorporate new information post-training, leading to three primary challenges. First, the focus on broad and general data to maximize accessibility and applicability results in subpar performance in specialized areas. Second, the rapid creation of online data, combined with the significant resources required for data annotation and model training, hinders LLMs' ability to stay updated. Third, LLMs are susceptible to generating convincing yet inaccurate responses, known as ``hallucinations'', which can mislead users.

\begin{figure}[t]
	\includegraphics[width=0.6\textwidth]{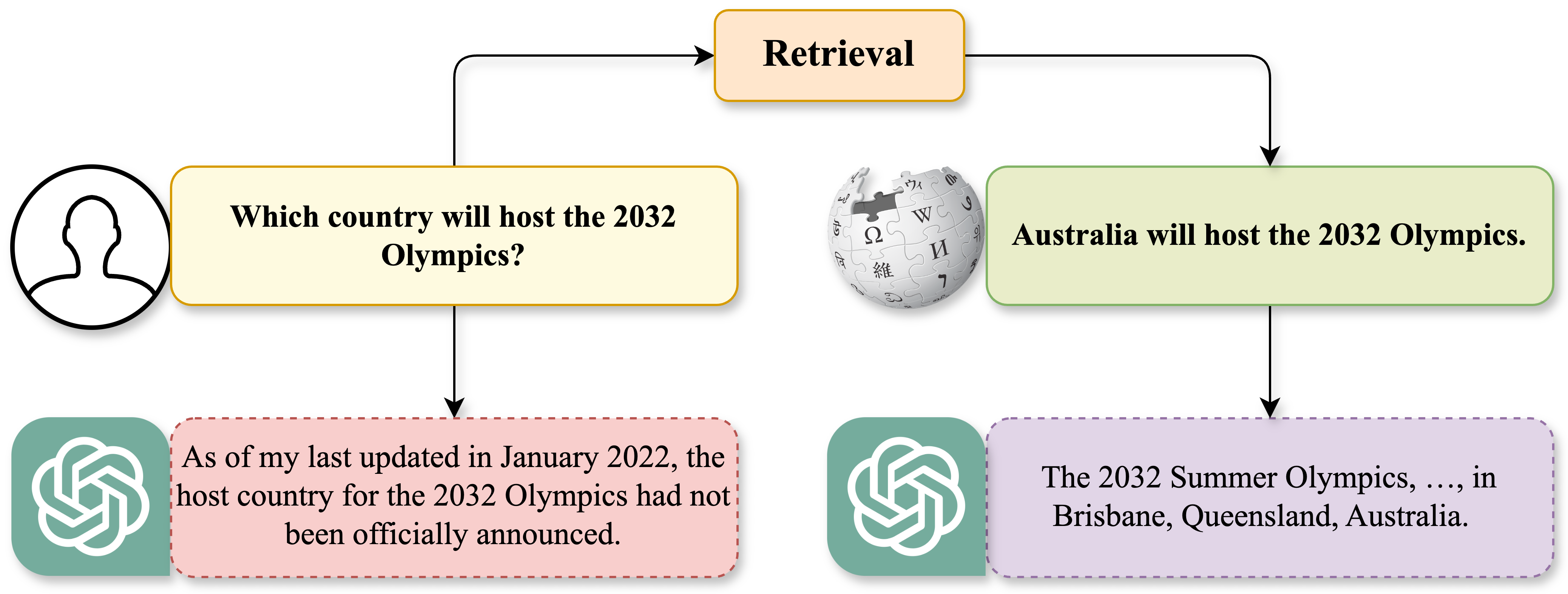}
	\caption{An example of RAG benefits ChatGPT resolves questions that cannot be answered beyond the scope of the training data and generates correct results.}
	\label{fig:ragexample}
\end{figure}

Addressing these challenges is crucial for LLMs to be effectively utilized across various domains. A promising solution is the integration of Retrieval-Augmented Generation (RAG) technology, which supplements models by fetching external data in response to queries, thus ensuring more accurate and current outputs. Figure~\ref{fig:ragexample} illustrates how RAG can enable ChatGPT to provide precise answers beyond its initial training data.

Since its introduction by Lewis et al. \cite{lewis2020retrievalaugmented} in 2020, RAG has seen rapid development, especially with the rise of models like ChatGPT. Despite these advancements, there remains a noticeable gap in the literature regarding a comprehensive analysis of the mechanisms underlying RAG and the progress achieved by subsequent studies. Moreover, the field suffers from fragmented research focuses and inconsistent terminology for similar methods, leading to confusion. This survey seeks to bridge this gap by offering a structured overview of RAG, categorizing various approaches, and providing an in-depth understanding of the current research landscape, with a focus on textual applications given their prominence in recent research.

To provide clarity and structure, this paper is organized as follows: Section 2 outlines the overall RAG workflow, dividing the methodologies into pre-retrieval, retrieval, post-retrieval, and generation phases. Sections 3 through 6 explore the core techniques within each phase. Section 7 focuses on the evaluation methodologies for RAG. Section 8 summarizes the reviewed studies, detailing the retrievers and generators used, while Section 9 discusses challenges and future research directions, extending beyond text-based studies to include multimodal data applications. The paper concludes with Section 10.

Other related surveys provide valuable insights into the evolving RAG landscape from different angles. Gao et al. \cite{gao2023retrievalaugmented} identified three key stages in RAG development: pre-training enhancement, inference, and fine-tuning. Zhao et al. \cite{zhao2024retrievalaugmented} focused on the diverse applications of RAG, including text, code, image, and video generation, emphasizing augmented intelligence in generative tasks. Meanwhile, Hu et al. \cite{hu2024rau} explored Retrieval-Augmented Language Models (RALMs), examining how interactions between retrievers, language models, and augmentations influence model architectures and applications.

In this paper, we aim to offer a comprehensive and unified framework for understanding RAG from an information retrieval (IR) perspective, identifying key challenges and areas for improvement. We delve into the core technologies that drive RAG, assessing their effectiveness in addressing retrieval and generation tasks. Additionally, this survey introduces the evaluation methods employed in RAG research, highlights current limitations, and proposes promising avenues for future exploration.

\section{RAG Framework}
\label{sec:framework}
\begin{figure}
	\includegraphics[width=\textwidth]{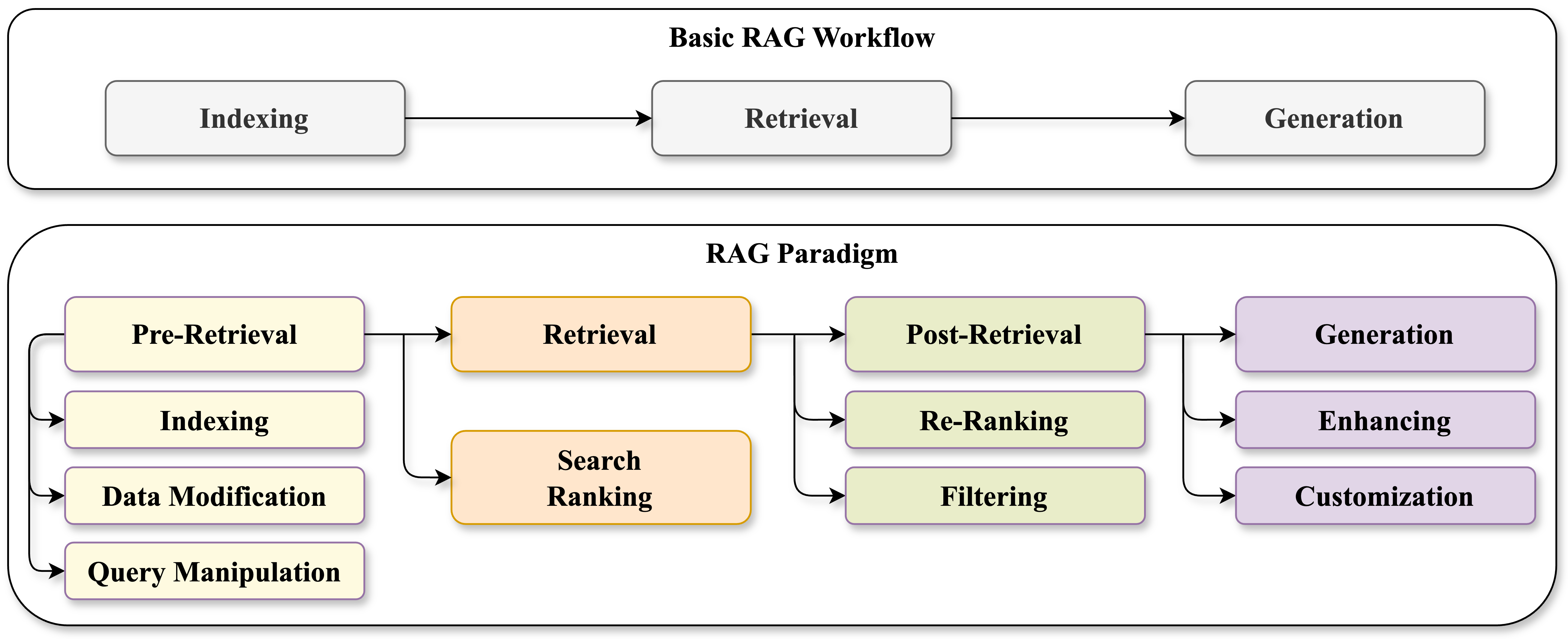}
	\caption{The unified RAG core concepts with basic workflow.}
	\label{fig:rag_paradigm}
\end{figure}

The hallucinations are largely attributed to LLMs' inability to access up-to-date information. This limitation stems from the models' reliance on their training datasets. RAG proposes a solution to this issue by supplementing the LLM's training data with current information from external sources through a retrieval model, thereby enabling the generation of accurate responses. RAG presents a more cost-effective alternative to the extensive training and fine-tuning processes typically required for LLMs. It allows for the dynamic incorporation of fresh information via traditional retrieval methods or pre-trained LMs, without the need to directly integrate this new data into the LLM. This feature makes RAG both flexible and scalable, facilitating its application across different LLMs for various purposes. The information retrieved through RAG is derived from real-world data, authored by humans, which not only simplifies the generation process but also increases the reliability of the generated responses. 

Research by Khandelwal et al. \cite{khandelwal2020generalization} demonstrates that accessing relevant information from the training dataset itself can significantly improve LLM performance, highlighting the effectiveness of RAG. Over time, RAG has evolved from a means of providing supplementary information to enabling multiple interactions between the retrieval and generation components. This involves conducting several rounds of retrieval to refine the accuracy of the information retrieved and iteratively improve the quality of the generated output. Toolkits such as LangChain\footnote{https://www.langchain.com} and LlamaIndex\footnote{https://www.llamaindex.ai} have modularized the RAG approach, enhancing its adaptability and expanding its range of applications. Despite these toolkits employing diverse methodologies to tackle different aspects of RAG—from multiple search iterations to iterative generation—they maintain adherence to the fundamental RAG workflow. This consistency is crucial for understanding their operation and pinpointing opportunities for further development.

\subsection{Basic RAG Workflow}
 Figure \ref{fig:rag_paradigm} represents the unified RAG core concepts with basic workflow. The workflow of RAG begins with the creation of an index comprising external sources. This index serves as the basis for retrieving relevant information through a retriever model based on a specific query. The final step involves a generator model, which combines the retrieved information with the query to produce the desired output. 

\subsubsection{Indexing}
Efficient retrieval begins with comprehensive indexing, where data preparation is key. This stage involves text normalization processes such as tokenization, stemming, and the removal of stop words to enhance the text's suitability for indexing \cite{DBLP:books/daglib/0021593}. Text segments are then organized into sentences or paragraphs to facilitate more focused searches, allowing for the pinpointing of segments containing pertinent keywords. The integration of deep learning has revolutionized indexing through the use of pretrained LMs for generating semantic vector representations of texts. These vectors are stored, enabling rapid and precise retrieval from extensive data collections, significantly enhancing retrieval efficiency.

\subsubsection{Retrieval}
While traditional retrieval methods, such as the BM25 algorithm \cite{DBLP:conf/trec/Hancock-BeaulieuGHRWW96}, focus on term frequency and presence for document ranking, they often overlook the semantic information of queries. Current strategies leverage pretrained LMs like BERT \cite{devlin2019bert}, which capture the semantic essence of queries more effectively. These models improve search accuracy by considering synonyms and the structure of phrases, thereby refining document ranking through the detection of semantic similarities. This is typically achieved by measuring vector distances between documents and queries, combining traditional retrieval metrics with semantic understanding to yield search results that are both relevant and aligned with user intent.

\subsubsection{Generation}
The generation phase is tasked with producing text that is both relevant to the query and reflective of the information found in the retrieved documents. The usual method involves concatenating the query with the retrieved information, which is then fed into an LLM for text generation \cite{li2022survey}. Although ensuring the generated text's alignment and accuracy with the retrieved content presents challenges, it is also essential to strike a balance between adhering closely to the source material and infusing the output with creativity. The generated text should accurately convey the information from the retrieved documents and align with the query's intent, while also offering the flexibility to introduce new insights or perspectives not explicitly contained within the retrieved data.

\subsection{RAG Paradigm}
The RAG paradigm organizes research within the domain, offering a straightforward yet robust framework to enhance LLM performance. Central to RAG is its search mechanism, crucial for generating high-quality outcomes. Therefore, this paradigm is structured into four main phases from a retrieval perspective: pre-retrieval, retrieval, post-retrieval, and generation. Both single-hop and multi-hop retrieval approaches, encompassing iterative retrieve-generate cycles, follow this four-phase structure. Figure \ref{fig:rag tax} is the taxonomy tree of RAG's core techniques.

\subsubsection{Pre-Retrieval}
The pre-retrieval phase of retrieval-augmented generation lays the foundation for successful data and query preparation, ensuring efficient information retrieval. This phase includes essential tasks to prepare for effective data access.

\paragraph{Indexing} The process starts with indexing, which establishes an organized system to enable fast and accurate retrieval of information. The specificity of indexing depends on the task and data type. For example, sentence-level indexing is beneficial for question-answering systems to precisely locate answers, while document-level indexing is more appropriate for summarizing documents to understand their main concepts and ideas.

\paragraph{Query Manipulation} After indexing, query manipulation is performed to adjust user queries for a better match with the indexed data. This involves query reformulation \cite{DBLP:journals/jasis/JansenBS09, DBLP:conf/sigir/YuLYXBG020}, which rewrites the query to align more closely with the user's intention; query expansion \cite{DBLP:journals/ipm/HuangMH13}, which extends the query to capture more relevant results through synonyms or related terms; and query normalization, which resolves differences in spelling or terminology for consistent query matching.

\paragraph{Data Modification} Data modification is also critical in enhancing retrieval efficiency. This step includes preprocessing techniques like removing irrelevant or redundant information to improve the quality of results and enriching the data with additional information such as metadata to boost the relevance and diversity of the retrieved content \cite{DBLP:conf/nips/BevilacquaOLY0P22}.

\subsubsection{Retrieval}
\paragraph{Search \& Ranking} The retrieval stage is the combination of search and ranking. It focuses on selecting and prioritizing documents from a dataset to enhance the quality of the generation model's outputs. This stage employs search algorithms to navigate through the indexed data, finding documents that match a user's query.  After identifying relevant documents, the process of initially ranking these documents starts to sort them according to their relevance to the query. 

\subsubsection{Post-Retrieval}
The post-retrieval phase serves to refine the initially retrieved documents to improve the quality of text generation. This phase consists of re-ranking and filtering, each aimed at optimizing the document selection for the final generation task.

\paragraph{Re-Ranking}
In the re-ranking step, the documents previously retrieved are reassessed, scored, and reorganized. The objective is to more accurately highlight the documents most relevant to the query and diminish the importance of the less relevant ones. This step involves incorporating additional metrics and external knowledge sources to enhance precision. In this context, pre-trained models with superior accuracy but lower efficiency can be effectively employed due to the limited set of candidate documents available \cite{DBLP:conf/sigir/HuangH09}. 

\paragraph{Filtering} Filtering aims to remove documents that fail to meet specified quality or relevance standards \cite{khattab2020colbert, DBLP:conf/ecai/HuangH23}. This can be done through several approaches, such as establishing a minimum relevance score threshold to exclude documents below a certain relevance level. Furthermore, the use of feedback from users or prior relevance evaluations assists in adjusting the filtering process, guaranteeing that only the most relevant documents are retained for text generation.

\subsubsection{Generation}
The generation stage is a crucial component of the RAG process, responsible for leveraging retrieved information to enhance the quality of the generated response. This stage encompasses several sub-steps aimed at producing content that is readable, engaging, and informative.

\paragraph{Enhancing}
At the heart of the generation phase is the enhancement step, where the objective is to merge the retrieved information with the user's query to create a coherent and relevant response. This includes the process of elaboration, adding extra details to the retrieved content to enrich it. Efforts are focused on improving the output's quality by increasing its clarity, coherence, and stylistic appeal through methods such as rephrasing and restructuring. Information from various sources is combined to offer a comprehensive perspective, and verification is conducted to ensure the accuracy and relevance of the content.

\paragraph{Customization}
Customization is user-centric. It encompasses tailoring content in two primary ways. First, it aligns the generated output with relevant information retrieved in earlier stages, ensuring consistency and accuracy by incorporating key knowledge. Second, it adapts the content to suit user-specific factors such as intended audience, situational context, and personal preferences, shaping the response to be both contextually relevant and user-centric. This dual focus on integrating relevant knowledge and adjusting to diverse contextual demands forms the basis of effective customization in RAG.

\section{Pre-Retrieval}
\label{sec:preretrieval}
\begin{figure}[t]
	\centering
	\tikzset{
		root/.style   = {align=center, fill=f_beige, text width=6cm, font=\Huge},
		xnode/.style = {align=center, fill=blue!15, text width=6cm, font=\Huge},
		tnode/.style = {align=center, fill=pink!50, text width=5cm, font=\huge},
		pnode/.style = {align=center, fill=pink!50, text width=5cm, font=\LARGE},
		edge from parent/.style={draw=black, very thick, edge from parent fork south},
	}
	\resizebox{\textwidth}{!}{%
		\begin{forest}
			for tree={
				grow=south,
				growth parent anchor=south,
				parent anchor=south,
				child anchor=north,
				draw,
				minimum height = 1.2cm,
				rounded corners=10pt,
				drop shadow,
				node options={align=center,},
				text width=3cm,
				l sep=10mm,
				s sep=3mm,
				edge={very thick, draw=black}, 
				tier/.wrap pgfmath arg={tier #1}{level()},
				edge path={
					\noexpand\path [\forestoption{edge}, very thick]
					(!u.parent anchor) -- +(0,-15pt) -|   
					(.child anchor)\forestoption{edge label};
				},
			}
			[RAG, root, for tree={parent anchor=south}
			[Pre-Retrieval, xnode, for tree={parent anchor=south}
			[Indexing, tnode, for tree={parent anchor=south}
			[{REALM \cite{guu2020retrieval}; \\ kNN-LMs \cite{khandelwal2020generalization}; \\RAG \cite{lewis2020retrievalaugmented}; \\Webgpt \cite{nakano2021webgpt}; \\RETRO \cite{borgeaud2022improving}; \\MEMWALKER \cite{chen2023walking}; \\Atlas \cite{ma2023query}; \\ Chameleon \cite{jiang2023chameleon}; \\ AiSAQ \cite{tatsuno2024aisaq}; \\PipeRAG \cite{jiang2024piperag}; \\LRUS-CoverTree \cite{ma2024reconsidering}
			}, pnode]]
			[Query Manipulation, tnode, for tree={parent anchor=south}
			[{Webgpt \cite{nakano2021webgpt}; \\DSP \cite{khattab2022demonstratesearchpredict};\\ CoK \cite{li2024chainofknowledge};\\ IRCOT \cite{trivedi2023interleaving};\\ Query2doc \cite{wang2023querydoc};\\ Step-Back \cite{zheng2024take};\\ PROMPTAGATOR \cite{dai2023promptagator};\\ KnowledGPT \cite{wang2023knowledgpt}; \\Rewrite-Retrieve-Read \cite{ma2023query}; \\FLARE \cite{jiang2023active}; \\RQ-RAG \cite{chan2024rqrag}; \\RARG \cite{yue2024evidencedriven}; \\DRAGIN \cite{su2024dragin}
			}, pnode]]
			[Data Modification, tnode, for tree={parent anchor=south}
			[{RA-DIT \cite{lin2024radit}; \\RECITE \cite{sun2023recitationaugmented}; \\UPRISE \cite{cheng2023uprise}; \\GENREAD \cite{yu2023generate}; \\KnowledGPT \cite{wang2023knowledgpt}; \\Selfmem \cite{cheng2023lift}; \\RARG \cite{yue2024evidencedriven}
			}, pnode] ]
			]
			[Retrieval, xnode, for tree={parent anchor=south}
			[Search \& Ranking, tnode, for tree={parent anchor=south}
			[{REALM \cite{guu2020retrieval}; \\kNN-LMs \cite{khandelwal2020generalization}; \\RAG \cite{lewis2020retrievalaugmented}; \\FiD \cite{izacard2021leveraging}; \\Webgpt \cite{nakano2021webgpt}; \\RETRO \cite{borgeaud2022improving}; \\ITRG \cite{feng2024retrievalgeneration}; \\RA-DIT \cite{lin2024radit}; \\SURGE \cite{kang2023knowledge}; \\PRCA \cite{yang2023prca}; \\AAR \cite{yu2023augmentationadapted}; \\ITER-RETGEN \cite{shao2023enhancing}; \\UPRISE \cite{cheng2023uprise}; \\MEMWALKER \cite{chen2023walking}; \\Atlas \cite{ma2023query}; \\FLARE \cite{jiang2023active}; \\PlanRAG \cite{lee2024planrag}
			}, pnode] ]]
			[Post-Retrieval, xnode, for tree={parent anchor=south}
			[Re-Ranking, tnode, for tree={parent anchor=south}
			[{Re2G \cite{glass2022reg}; \\DSP \cite{khattab2022demonstratesearchpredict}; \\CoK \cite{li2024chainofknowledge}; \\FiD-TF \cite{berchansky2023optimizing}; \\ITER-RETGEN \cite{shao2023enhancing}; \\PROMPTAGATOR \cite{dai2023promptagator}; \\Selfmem \cite{cheng2023lift}; \\DKS-RAC \cite{huang2023retrieval}; \\In-Context RALM \cite{ram2023incontext}; \\Fid-light \cite{hofstätter2023fidlight}; \\GenRT \cite{xu2024listaware}
			}, pnode]]
			[Filtering, tnode, for tree={parent anchor=south}
			[{Webgpt \cite{nakano2021webgpt}; \\Self-RAG \cite{asai2024selfrag}; \\FiD-TF \cite{berchansky2023optimizing}; \\PROMPTAGATOR \cite{dai2023promptagator}; \\RECOMP \cite{xu2024recomp}; \\DKS-RAC \cite{huang2023retrieval}; \\CoK \cite{li2024chainofknowledge}; \\FILCO \cite{wang2023learning}; \\BlendFilter \cite{wang2024blendfilter}; \\CRAG \cite{yan2024corrective}
			}, pnode] ] ] 
			[Generation, xnode, for tree={parent anchor=south}
			[Enhancing, tnode, for tree={parent anchor=south}
			[{FiD \cite{izacard2021leveraging}; \\Webgpt \cite{nakano2021webgpt}; \\DSP \cite{khattab2022demonstratesearchpredict}; \\IRCOT \cite{trivedi2023interleaving}; \\ITRG \cite{feng2024retrievalgeneration}; \\RA-DIT \cite{lin2024radit}; \\PRCA \cite{yang2023prca}; \\RECITE \cite{sun2023recitationaugmented}; \\UPRISE \cite{cheng2023uprise}; \\GENREAD \cite{yu2023generate}; \\Selfmem \cite{cheng2023lift}; \\MEMWALKER \cite{chen2023walking}; \\Atlas \cite{ma2023query}
			}, pnode]]
			[Customization, tnode, for tree={parent anchor=south}
			[{LAPDOG \cite{huang2023learning}; \\PersonaRAG \cite{zerhoudi2024personarag}; \\ERAGent \cite{shi2024eragent}; \\ROPG \cite{salemi2024optimization}
			}, pnode] ]] 
			]
		\end{forest}
	}
	\caption{Taxonomy tree of RAG’s core techniques}
	\label{fig:rag tax}
	
\end{figure}

\subsection{Indexing}
One of the most commonly used indexing structures in traditional information retrieval systems is the inverted index. This structure associates documents with words to form a vocabulary list, allowing users to quickly locate references where a specific word appears within a collection of documents. The vocabulary list here refers to the set of all unique words present in the document collection, while the reference includes the documents where the word appears, along with the word's position and weight within those documents. However, traditional indexing structures struggle to retrieve documents that are semantically related to a user's query but do not contain the exact query terms.

To address this limitation, retrieval methods using dense vectors generated by deep learning models have become the preferred choice. These vectors, also known as embeddings, capture the semantic meaning of words and documents, allowing for more flexible and accurate retrieval. Dense vector-based indexing methods can be categorized into three main types: graphs, product quantization (PQ) \cite{jégou2011product}, and locality-sensitive hashing (LSH) \cite{datar2004localitysensitive}. Since generating dense vectors with large language models requires substantial resources, and the document collections to be searched are typically vast, the core strategy of these indexing methods is based on approximate nearest neighbor search (ANNS) \cite{DBLP:journals/jacm/AryaMNSW98}. This approach significantly speeds up the search process at the cost of a slight reduction in search accuracy. 

\paragraph{Graph} Using graphs to build indexes is a common practice in RAG. By indexing vectors with a graph structure, the range of nodes where distances need to be computed during retrieval can be limited to a local subgraph, thereby enhancing search speed. Several prominent methods and tools have been developed using this approach. For example, k-nearest neighbor language models kNN-LMs \cite{khandelwal2020generalization}, as demonstrated by Khandelwal et al., integrate the kNN algorithm with pre-trained language models. This method employs a datastore created from collections of texts to dynamically retrieve contextually relevant examples, enhancing model performance without requiring additional training. FAISS \cite{8733051}, a tool widely adopted for indexing in many studies \cite{khandelwal2020generalization, lewis2020retrievalaugmented, khattab2022demonstratesearchpredict}, integrates enhancements like the Hierarchical Navigable Small World (HNSW) approximation \cite{8594636} to further speed up retrieval \cite{lewis2020retrievalaugmented}. WebGPT \cite{nakano2021webgpt} showcases another practical application by utilizing the Bing API\footnote{https://www.microsoft.com/en-us/bing/apis/bing-web-search-api} for indexing based on actual user search histories, which illustrates the potential of integrating real-world user data into the retrieval process. Additionally, other methods like MEMWALKER \cite{chen2023walking} introduces innovative approaches to overcome limitations such as context window size in large language models. It creates a memory tree from input text, segmenting the text into smaller pieces and summarizing these segments into a hierarchical structure. Moreover, LRUS-CoverTree method \cite{ma2024reconsidering} designed another tree structure for k-Maximum Inner-Product Search (k-MIPS) and achieves performance comparable with significantly lower index construction time. These techniques facilitate efficient indexing and management of large information volumes, demonstrating the versatility and effectiveness of graph-based approaches.

\paragraph{Product Quantization} PQ is one of the most representative methods for handling large-scale data. It accelerates searches by segmenting vectors and then clustering each part for quantization. Unlike graph-based methods, which speed up searches by reducing the number of vectors for distance calculation, PQ achieves faster searches by reducing the time spent on calculating word distances. Several implementations of PQ have emerged in RAG, each improving its efficiency and scalability in different ways. PipeRAG \cite{jiang2024piperag} integrates PQ within a pipeline-parallelism framework to enhance retrieval-augmented generation by optimizing retrieval intervals. Chameleon system \cite{jiang2023chameleon} leverages PQ in a disaggregated accelerator environment to balance memory usage and retrieval speed in RAG tasks. AiSAQ \cite{tatsuno2024aisaq} introduces an all-in-storage ANNS method that offloads PQ vectors from DRAM to storage, drastically reducing memory usage while maintaining high recall. It demonstrates that even with billion-scale datasets, memory usage can be minimized to around 10 MB with only minor latency increases, making it a highly scalable solution for RAG systems. 

\paragraph{Locality-sensitive Hashing} The core idea of LSH is to place similar vectors into the same hash bucket with high probability. LSH uses hash functions that map similar vectors to the same or nearby hash values, making it easier to find approximate nearest neighbours. In LSH, when a query vector is hashed, the system quickly retrieves candidate vectors that share the same hash value. This method reduces the dimensionality of the problem and can be implemented efficiently, but it may introduce some inaccuracies due to the hashing process itself. While LSH is rarely used in RAG systems compared to graph-based and PQ methods, it still offers a useful approach in scenarios where speed is prioritized over the slight loss in accuracy.

\subsection{Query Manipulation}
Query manipulation is pivotal in enhancing the effectiveness and accuracy of modern IR systems. By refining users' original queries, it addresses challenges such as ambiguous phrasing and vocabulary mismatches between the query and target documents. This process involves more than merely replacing words with synonyms; it requires a deep understanding of user intent and the context of the query, particularly in complex tasks like RAG. Effective query manipulation significantly boosts retrieval performance, which in turn can greatly impact the quality of generated outputs. The three primary approaches to query manipulation are query expansion, query reformulation, and prompt-based rewriting.

\paragraph{Query Expansion} Query expansion involves augmenting the original query with additional terms or phrases that are related to or synonymous with the query terms. In the context of LLMs, query expansion can be more sophisticated, utilizing the model's extensive knowledge to generate contextually relevant expansions. This technique aims to improve recall by ensuring that the retrieval process captures a broader range of relevant documents, accommodating different terminologies or expressions. Techniques such as synonym expansion, semantic similarity, or leveraging external knowledge bases are commonly employed in query expansion. For example, the method described in FiD \cite{izacard2021leveraging} expands the query by retrieving a wider range of passages using both sparse and dense retrieval techniques, enabling the model to aggregate evidence from multiple sources and thereby improving the accuracy and robustness of generated answers. A more advanced form of query expansion is demonstrated in Query2doc \cite{wang2023querydoc}, where pseudo-documents generated by LLMs enhance the original query, effectively bridging the gap between the user's input and the corpus information, which benefits both sparse and dense retrieval systems. Similarly, KnowledGPT \cite{wang2023knowledgpt} broadens the scope of information accessed during retrieval by leveraging external knowledge bases, further refining the retrieval process. The RARG \cite{yue2024evidencedriven} framework uses an evidence-driven query expansion approach, incorporating a wide array of supporting documents to generate informed and accurate counter-misinformation responses.

\paragraph{Query Reformulation} Query reformulation involves rephrasing or restructuring the original query to enhance its effectiveness. This might include making the wording more specific, removing vague terms, or adjusting the syntax to better align with the retrieval system's requirements. With LLMs, query reformulation can be dynamically driven by understanding the user's intent and context, allowing for more precise modifications that lead to improved retrieval results. This reformulation process can also be informed by past queries or user interactions, adapting the query to better fit the specific retrieval task. For instance, the RQ-RAG \cite{chan2024rqrag} model represents an advanced form of query reformulation by rewriting, decomposing, and disambiguating queries, making it particularly effective in scenarios that demand complex query handling. This approach ensures that the refined query better matches the needed context, improving the relevance of retrieved information. Rewrite-Retrieve-Read framework \cite{ma2023query} adjusts the original query to optimize the retrieval process, allowing the system to more effectively leverage retrieved data for generating accurate responses. Additionally, FLARE \cite{jiang2023active} exemplifies query reformulation through its active retrieval-augmented generation approach, which iteratively refines the query based on a feedback loop between retrieval and generation, thereby enhancing the accuracy and relevance of the retrieved information.

\paragraph{Prompt-based Rewriting} Prompt-based rewriting, particularly in the context of LLMs, represents an innovative approach where the original query is embedded within a larger prompt or context to guide the LLM's response. This technique harnesses the model's ability to understand and generate language within a specific context, effectively rewriting the query to align with the desired output. Prompt-based rewriting is especially powerful in scenarios where the retrieval process is integrated into a generative workflow, allowing the system to adapt the query to various stages of retrieval and generation. This approach may also involve dynamic prompts that evolve based on interaction, further refining the retrieval process. For example, Step-Back \cite{zheng2024take} refines the query context through carefully crafted prompts that guide the LLM's reasoning process, ensuring that the outputs are more aligned with the user's intent, particularly in complex reasoning tasks. The CoK \cite{li2024chainofknowledge} method focuses on dynamically adapting the knowledge source and using prompts to rewrite the context in which a query is interpreted. This approach leverages prompt-based rewriting to enable the LLM to effectively integrate and ground its responses based on various heterogeneous knowledge sources. Additionally, Promptagator \cite{dai2023promptagator} discusses using prompt-based techniques to adapt and rewrite the query to better align with the retrieval system's expectations, particularly in few-shot learning scenarios. These prompts guide the model in generating or refining the query to optimize retrieval results.

\subsection{Data Modification}
Document modification techniques play a critical role in enhancing retrieval performance, particularly when integrated with LLMs. These techniques can be broadly categorized into Internal Data Augmentation and External Data Enrichment. Internal Data Augmentation focuses on maximizing the value of existing information within documents or models, while External Data Enrichment introduces supplementary data from outside sources to fill gaps, provide additional context, or broaden the scope of the content.

\paragraph{Internal Data Augmentation}
Internal Data Augmentation leverages information already present within documents or taps into the inherent knowledge embedded in LLMs. Techniques like paraphrasing, where content is rewritten for improved readability or multiple perspectives, and summarization, which condenses information while retaining core content, are commonly employed. Other methods involve generating supplementary content or explanations that are contextually related without introducing external data. For instance, RECITE \cite{sun2023recitationaugmented} utilizes a model’s internal memory to recite relevant information before generating responses, thus enhancing performance in tasks like closed-book question answering without external data. KnowledGPT \cite{wang2023knowledgpt} similarly refines the internal knowledge embedded within LLMs, optimizing its use during generation. GENREAD \cite{yu2023generate} further demonstrates how pre-existing knowledge within LLMs can be used to generate context that enhances task performance, bypassing the need for external sources. In another example, the Selfmem \cite{cheng2023lift} framework allows the model to iteratively use its own outputs as memory in subsequent generation tasks. By selecting and utilizing the best internal outputs as memory, this approach boosts model performance without depending on external memory resources.

\paragraph{External Data Enrichment}
External Data Enrichment enhances document content by incorporating new information from external sources, enriching the overall context and accuracy. This process can involve integrating facts, data, or contextual knowledge from external datasets or knowledge bases. For example, RA-DIT \cite{lin2024radit} augments input prompts during fine-tuning by leveraging large datasets like Wikipedia and CommonCrawl, enhancing the model’s capability in knowledge-intensive tasks. The dual instruction tuning technique optimizes both the language model and the retriever to more effectively incorporate retrieved information. UPRISE \cite{cheng2023uprise} demonstrates how retrieving prompts from diverse task datasets improves model generalization in zero-shot scenarios by enriching the context during inference. Additionally, RARG \cite{yue2024evidencedriven} exemplifies external data enrichment by integrating scientific evidence from academic databases to strengthen responses countering misinformation. This method involves a two-stage retrieval pipeline that identifies and ranks relevant documents, which are then used to support and enhance the factual accuracy of generated responses.

\section{Retrieval}
\label{sec:retrieval}
\begin{figure}
	\includegraphics[width=0.85\textwidth]{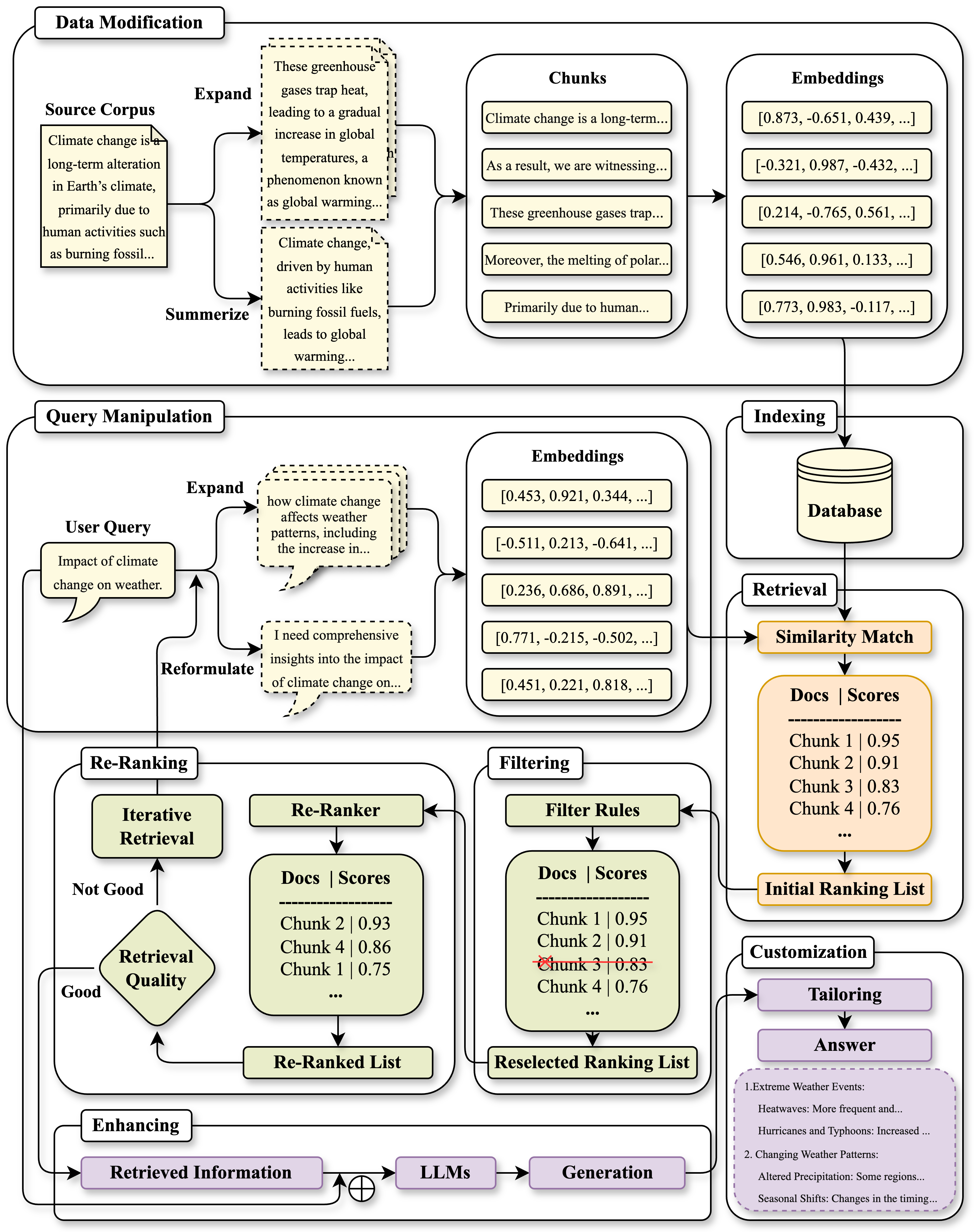}
	\caption{An example of a typical RAG framework with interative retrieval strategy.}
	\label{fig:rag_framework}
\end{figure}

\subsection{Search \& Ranking}
The search and ranking process within RAG is crucial for improving the relevance and accuracy of generated outputs. Several methodologies have been developed to refine this process, each contributing unique strategies for enhancing retrieval and ranking. For example, Atlas \cite{izacard2023atlas} and AAR \cite{yu2023augmentationadapted} both aim to improve the relevance of retrieved documents, but they approach this challenge differently. Atlas focuses on optimizing the retriever's ability to select contextually relevant documents, especially in new domains with limited data, by employing few-shot learning techniques such as Attention Distillation and Perplexity Distillation. AAR, on the other hand, adapts retrieval preferences to better align with the requirements of LLMs, enhancing retrieval generalization across tasks by training a smaller source model.

Additionally, IRCOT \cite{trivedi2023interleaving} and FLARE \cite{jiang2023active} introduce dynamic interactions within the retrieval process, albeit with distinct goals. IRCOT integrates retrieval with chain-of-thought (CoT) reasoning, interleaving these processes to ensure that each retrieval step supports the ongoing reasoning task. FLARE, in contrast, adopts a confidence-based active retrieval mechanism, dynamically triggering retrieval when the model generates low-confidence tokens. This approach is particularly useful in scenarios where model confidence varies, as it allows the system to fetch additional information to resolve uncertainties during the generation process.

When addressing domain-specific retrieval challenges, SURGE \cite{kang2023knowledge} and PRCA \cite{yang2023prca} offer different solutions. SURGE uses a subgraph retriever to extract relevant subgraphs from knowledge graphs, integrating structured data into the retrieval process to improve the contextual understanding of generated responses. The relational structure of knowledge graphs allows for more accurate and informed retrieval. PRCA, in contrast, focuses on domain-specific abstractive summarization, using a reward-driven approach to refine the retrieved content. This strategy is designed to optimize content for the generator, particularly in scenarios where the generator functions as a black box, thereby enhancing alignment between retrieval and generation.

MEMWALKER \cite{chen2023walking} presents a unique approach to handling long-context question answering by incorporating an internal search and ranking mechanism within a memory tree structure. This method navigates extensive memory stores, ensuring that the most relevant information is retrieved and used for complex queries. Unlike other methods, MEMWALKER emphasizes efficient processing of long texts through iterative navigation and summarization, rather than solely optimizing the initial retrieval phase.

\subsection{Retrieval Strategy}
The retrieval strategies within RAG are vital for customizing the retrieval process to specific application needs, with each strategy offering distinct advantages and addressing particular challenges. In RAG, it is mostly the utilization of retrieval techniques rather than the exploration of retrieval algorithms that is involved, so it is the strategy of retrieval that is usually considered in searching and ranking. While basic RAGs are usually single-hop searches, i.e., they are retrieved only once as generated supplementary material, today's RAGs are mostly multi-hop searches, i.e., they are searched several times through different search strategies until they are satisfied. In terms of practical applications these strategies belong to the design on the engineering pipeline. Figure \ref{fig:rag_framework} shows a typical case of the RAG framework with iterative retrieval strategy. There are five main retrieval strategies in RAG:

\paragraph{Basic Retrieval Strategy} Basic retrieval strategies typically follow a linear workflow, moving sequentially through pre-retrieval, retrieval, post-retrieval, and generation phases. The Atlas \cite{ma2023query} framework exemplifies this straightforward approach, guiding the retrieval process efficiently from start to finish without iterations or complex conditional modifications. REPLUG \cite{shi2023replug} similarly follows this basic strategy, augmenting black-box language models with retrieval in a simple manner, where the retrieved information is directly used to enhance the generation process.

\paragraph{Iterative Retrieval Strategy} For more complex scenarios, iterative retrieval strategies (Algorithm \ref{alg:iterative_rag_workflow}) are employed, where information is retrieved in multiple steps, each informed by previous results. IRCOT \cite{trivedi2023interleaving} exemplifies this by integrating retrieval with chain-of-thought reasoning, where the retrieval process is sequential and closely tied to reasoning steps. This method is particularly effective in scenarios requiring multi-step problem-solving, such as research assistance or complex queries that benefit from detailed exploration. ITER-RETGEN \cite{shao2023enhancing} also employs iterative retrieval, refining the process based on generated responses, allowing for continuous improvement and closer alignment between retrieval and generation. RQ-RAG \cite{chan2024rqrag} advances this approach by using techniques like query rewriting, decomposition, and disambiguation, refining the retrieval step-by-step to enhance the final output. PlanRAG \cite{lee2024planrag} also fits within this strategy, iteratively refining the retrieval process based on generated content and feedback, ensuring that each step is better informed than the last.

\begin{algorithm}
	\caption{Iterative Retrieval Strategy in RAG}
	\label{alg:iterative_rag_workflow}
	\begin{algorithmic}[1]
		\REQUIRE Query $q$, Documents $D$, Maximum Iterations $N$, Retriever $R$, Generator $G$, Pre-retrieval Function $F_{pre}$, Post-retrieval Function $F_{post}$
		\ENSURE Final Output $y_{final}$
		\STATE Initialize $i \gets 1$ \hfill \textit{// Start iteration counter}
		
		\WHILE{$i \leq N$}
		\STATE \textbf{Pre-retrieval Phase}
		\STATE $q' \gets F_{pre}(q)$ \hfill \textit{// Indexing, Query Manipulation, Data Modification}
		
		\STATE \textbf{Retrieval Phase}
		\STATE $D_{i} \gets R(q', D)$ \hfill \textit{// Search and initial ranking of documents}
		
		\STATE \textbf{Post-retrieval Phase}
		\STATE $D_{i}' \gets F_{post}(q', D_{i})$ \hfill \textit{// Re-ranking and filtering to refine documents}
		
		\STATE \textbf{Generation Phase}
		\STATE $y_{i} \gets G(q', D_{i}')$ \hfill \textit{// Generate output based on refined documents}
		
		\IF{stopping condition met based on $y_{i}$}
		\STATE \textbf{BREAK} \hfill \textit{// Stop iterations if output is satisfactory}
		\ENDIF
		
		\STATE Update $q' \gets \text{UpdateQuery}(q, y_{i})$ \hfill \textit{// Refine query based on the generated output}
		\STATE $i \gets i + 1$ \hfill \textit{// Increment iteration counter}
		\ENDWHILE
		
		\STATE \textbf{Final Synthesis}
		\STATE $y_{final} \gets \text{SynthesizeResults}(\{y_{1}, y_{2}, \dots, y_{i}\})$ \hfill \textit{// Merge results}
		\RETURN $y_{final}$
		
	\end{algorithmic}
\end{algorithm}

\paragraph{Recursive Retrieval Strategy} Recursive retrieval (Algorithm \ref{alg:recursive_rag_workflow}) involves retrieval that can call itself, creating a hierarchy or tree of retrievals. This method effectively handles hierarchical or layered information by breaking down complex queries into simpler sub-queries. It is particularly useful for hierarchical data exploration, knowledge base construction, and detailed information retrieval. SURGE \cite{kang2023knowledge} leverages this strategy through knowledge graphs, where relevant subgraphs are extracted to enhance contextual understanding. The relational structure of knowledge graphs facilitates navigating multiple layers of information, ensuring accurate and contextually relevant retrieval. MEMWALKER \cite{chen2023walking} similarly adopts a recursive approach, processing long texts by constructing a memory tree of summaries. The system navigates through this tree to retrieve relevant information, effectively breaking down complex queries into manageable segments, which is particularly useful for handling long-context question answering. IMRAG \cite{yang2024imrag} introduces a multi-round retrieval mechanism, where each round of retrieval is based on the model's internal monologues, progressively refining the search with each iteration. Selfmem \cite{cheng2023lift} employs a self-memory module, enabling the system to store and retrieve information recursively, building upon previously retrieved knowledge in a hierarchical manner. This recursive strategy enhances the system’s ability to manage and integrate vast amounts of information across multiple retrieval iterations.

\begin{algorithm}
	\caption{Recursive Retrieval Strategy in RAG}
	\label{alg:recursive_rag_workflow}
	\begin{algorithmic}[1]
		\REQUIRE Initial Query $q$, Documents $D$, Maximum Depth $L$, Retriever $R$, Generator $G$, Pre-retrieval Function $F_{pre}$, Post-retrieval Function $F_{post}$, Sub-query Generation Function $F_{subq}$, Hierarchical Layer Building Function $F_{build}$, Hierarchical Information Operating Function $F_{hier}$
		\ENSURE Final Output $y_{final}$
		
		\STATE \textbf{Build Hierarchical Layers (Pre-retrieval)}
		\STATE $Hierarchy \gets F_{build}(q)$ \hfill \textit{// Build hierarchical layers based on the initial query}
		\STATE Initialize $l \gets 0$ \hfill \textit{// Start depth counter from 0}
		\STATE Initialize $Sub\_queries \gets [q]$ \hfill \textit{// Initialize list of sub-queries}
		
		\WHILE{$l \leq L$} 
		\FOR{each $q'_l \in Sub\_queries$}
		
		\STATE \textbf{Pre-retrieval Phase}
		\STATE $q'_l \gets F_{hier}(Hierarchy, q'_l)$ \hfill \textit{// Adjust query based on hierarchical layers}
		\STATE $q'_l \gets F_{pre}(q'_l)$ \hfill \textit{// Query Manipulation, Indexing, and Data Modification}
		
		\STATE \textbf{Retrieval Phase}
		\STATE $D_{l} \gets R(q'_l, D)$ \hfill \textit{// Retrieve documents for the current sub-query}
		
		\STATE \textbf{Post-retrieval Phase}
		\STATE $D_{l}' \gets F_{post}(q'_l, D_{l})$ \hfill \textit{// Re-ranking and filtering to refine documents}
		
		\STATE \textbf{Generation Phase}
		\STATE $y_{l} \gets G(q'_l, D_{l}')$ \hfill \textit{// Generate output based on refined documents}
		
		\STATE \textbf{Sub-query Generation (if needed)}
		\IF{additional refinement needed based on $y_{l}$}
		\STATE $Sub\_queries \gets F_{subq}(y_{l})$ \hfill \textit{// Generate new sub-queries based on current output}
		\ENDIF
		\ENDFOR
		
		\STATE $l \gets l + 1$ \hfill \textit{// Increment depth counter}
		\ENDWHILE
		
		\STATE \textbf{Final Synthesis}
		\STATE $y_{final} \gets \text{SynthesizeResults}(\{y_{0}, y_{1}, \dots, y_{l}\})$ \hfill \textit{// Merge results}
		\RETURN $y_{final}$
		
	\end{algorithmic}
\end{algorithm}

\paragraph{Conditional Retrieval Strategy} Conditional retrieval strategies (Algorithm \ref{alg:conditional_rag_workflow}) are governed by specific conditions or rules, which may be predefined or dynamically determined during the process. This method ensures that retrieval aligns with specific constraints or criteria, enhancing relevance and specificity. It is particularly useful for compliance checking, rule-based recommendation systems, and context-sensitive information retrieval. PRCA \cite{yang2023prca} is a prime example, where retrieval strategies are adapted based on reward-driven adjustments, refining the context used by large language models to enhance precision and relevance. RARG \cite{yue2024evidencedriven} similarly emphasizes retrieval based on specific evidence conditions, ensuring that the retrieval process aligns with predefined requirements, which is critical for generating factual and polite responses. CRAG \cite{yan2024corrective} adds another layer to this approach by incorporating a retrieval evaluator that assesses the quality of retrieved documents and triggers different actions based on confidence thresholds, ensuring that only the most relevant and accurate information is used in the generation process.

\begin{algorithm}
	\caption{Conditional Retrieval Strategy in RAG}
	\label{alg:conditional_rag_workflow}
	\begin{algorithmic}[1]
		\REQUIRE Query $q$, Documents $D$, Maximum Iterations $N$, Retriever $R$, Generator $G$, Pre-retrieval Function $F_{pre}$, Post-retrieval Function $F_{post}$, Condition Evaluation Function $F_{cond}$
		\ENSURE Final Output $y_{final}$
		
		\STATE Initialize $i \gets 1$ \hfill \textit{// Start iteration counter}
		\STATE $q' \gets q$ \hfill \textit{// Initialize query}
		
		\WHILE{$i \leq N$}
		
		\STATE \textbf{Pre-retrieval Phase}
		\STATE $q' \gets F_{pre}(q')$ \hfill \textit{// Perform query manipulation and data modification}
		
		\STATE \textbf{Retrieval Phase}
		\STATE $D_{i} \gets R(q', D)$ \hfill \textit{// Retrieve documents based on the current query}
		
		\STATE \textbf{Post-retrieval Phase}
		\STATE $D_{i}' \gets F_{post}(q', D_{i})$ \hfill \textit{// Re-rank and filter documents based on conditions}
		
		\STATE \textbf{Generation Phase}
		\STATE $y_{i} \gets G(q', D_{i}')$ \hfill \textit{// Generate output using the refined documents}
		
		\STATE \textbf{Conditional Branching}
		\IF{$F_{cond}(y_{i}, D_{i}')$ is Condition A}
		\STATE Apply Strategy A \hfill \textit{// e.g., refine the query based on feedback}
		\ELSIF{$F_{cond}(y_{i}, D_{i}')$ is Condition B}
		\STATE Apply Strategy B \hfill \textit{// e.g., expand the scope or adjust parameters}
		\ELSIF{$F_{cond}(y_{i}, D_{i}')$ is Condition C}
		\STATE Apply Strategy C \hfill \textit{// e.g., modify retrieval strategy or output processing}
		\ELSE
		\STATE Continue without changes \hfill \textit{// If no conditions are met, proceed without adjustments}
		\ENDIF
		
		\STATE \textbf{Check Termination Condition}
		\IF{$F_{cond}(y_{i}, D_{i}')$ meets stopping criteria}
		\STATE \textbf{BREAK} \hfill \textit{// Exit the loop if the stopping condition is met}
		\ENDIF
		
		\STATE $i \gets i + 1$ \hfill \textit{// Increment iteration counter}
		\ENDWHILE
		
		\STATE \textbf{Final Synthesis}
		\STATE $y_{final} \gets \text{SynthesizeResults}(\{y_{1}, y_{2}, \dots, y_{i}\})$ \hfill \textit{// Merge results}
		\RETURN $y_{final}$
		
	\end{algorithmic}
\end{algorithm}

\paragraph{Adaptive Retrieval Strategy} Adaptive retrieval (Algorithm \ref{alg:adaptive_rag_workflow}) dynamically adjusts the retrieval strategy based on the context and nature of the query or the data retrieved so far. This highly flexible method tailors retrieval approaches on-the-fly to optimize for relevance and precision, making it ideal for personalized search engines, adaptive learning systems, and real-time decision support. AAR  \cite{yu2023augmentationadapted} exemplifies adaptive retrieval by adjusting its strategy based on the preferences of LLMs, learning from a small source model and generalizing to unseen tasks. FLARE \cite{jiang2023active} takes a similar adaptive approach but focuses on dynamically fetching additional information when model confidence is low, thereby improving the relevance of generated responses. SelfRAG \cite{asai2024selfrag} goes further by incorporating self-reflective processes, where the retrieval strategy evolves based on critiques of the generated content. CoK \cite{li2024chainofknowledge}, on the other hand, implements a dynamic mechanism that adjusts retrieval strategies based on the evolving needs of the task. The retrieval process in CoK is not static but adapts according to the specific scenario and the nature of the information being accessed, making it highly effective for context-sensitive applications. DRAGIN \cite{su2024dragin} discusses a real-time dynamic retrieval mechanism that adapts to the evolving needs of the language model, ensuring that the retrieval strategy remains responsive and aligned with the immediate task requirements, thus optimizing the relevance and precision of the retrieved information. \\

\begin{algorithm}
	\caption{Adaptive Retrieval Strategy in RAG}
	\label{alg:adaptive_rag_workflow}
	\begin{algorithmic}[1]
		\REQUIRE Query $q$, Documents $D$, Maximum Iterations $N$, Retriever $R$, Generator $G$, Pre-retrieval Function $F_{pre}$, Post-retrieval Function $F_{post}$, Adaptive Adjustment Function $F_{adapt}$, Feedback Function $F_{feedback}$
		\ENSURE Final Output $y_{final}$
		
		\STATE Initialize $i \gets 1$ \hfill \textit{// Start iteration counter}
		\STATE $q', Context \gets q, \emptyset$ \hfill \textit{// Initialize query and context}
		
		\WHILE{$i \leq N$}
		
		\STATE \textbf{Pre-retrieval Phase}
		\STATE $q', Context \gets F_{pre}(q', Context)$ \hfill \textit{// Query Manipulation, Indexing, Data Modification, and Context Setup}
		
		\STATE \textbf{Dynamic Retrieval Phase}
		\STATE $D_{i} \gets R(q', D, Context)$ \hfill \textit{// Retrieve documents based on the current query and context}
		
		\STATE \textbf{Adaptive Post-retrieval Phase}
		\STATE $D_{i}' \gets F_{post}(q', D_{i}, Context)$ \hfill \textit{// Re-ranking and filtering based on adaptive criteria}
		
		\STATE \textbf{Generation Phase}
		\STATE $y_{i} \gets G(q', D_{i}', Context)$ \hfill \textit{// Generate output using the refined documents}
		
		\STATE \textbf{Adaptive Adjustment}
		\IF{$F_{feedback}(y_{i}, D_{i}')$ is negative}
		\STATE $q', Context \gets F_{adapt}(q', Context, y_{i}, D_{i}')$ \hfill \textit{// Dynamically adjust the query and context}
		\ENDIF
		
		\STATE \textbf{Feedback Integration}
		\IF{$F_{feedback}(y_{i})$ is positive}
		\STATE \textbf{BREAK} \hfill \textit{// Stop iterations if output is satisfactory}
		\ENDIF
		
		\STATE $i \gets i + 1$ \hfill \textit{// Increment iteration counter}
		\ENDWHILE
		
		\STATE \textbf{Final Synthesis}
		\STATE $y_{final} \gets \text{SynthesizeResults}(\{y_{1}, y_{2}, \dots, y_{i}\})$ \hfill \textit{// Merge results}
		\RETURN $y_{final}$
		
	\end{algorithmic}
\end{algorithm}

In summary, the choice of retrieval strategy within RAG depends on the specific requirements of the application at hand. While basic retrieval strategies offer simplicity and efficiency, iterative retrieval is well-suited for tasks requiring detailed exploration and refinement. Recursive retrieval excels in managing hierarchical information, while adaptive retrieval provides flexibility in dynamic environments. Conditional retrieval ensures strict adherence to predefined criteria, making it indispensable in applications where compliance and specific constraints are critical. By carefully selecting and combining these strategies, RAG systems can be tailored to effectively handle a wide range of information retrieval scenarios, leveraging the strengths of each approach to deliver robust and precise results.

\section{Post-Retrieval}
\label{sec:postretrieval}
\subsection{Re-Ranking}
As retrieval mechanisms often return a large number of potentially relevant documents, re-ranking methods are employed to reorder these documents, prioritizing those most likely to contribute meaningfully to the final output. By leveraging various strategies, including unsupervised techniques, supervised learning, and data augmentation, re-ranking aims to optimize the alignment between the retrieved content and the desired response, thereby improving the overall effectiveness of RAG systems \cite{zhu2023large}.

\paragraph{Unsupervised Re-ranking} Unsupervised re-rankers do not rely on labeled data for training. They use strategies such as pointwise, listwise, or pairwise methods to rank documents based on LLM outputs without the need for supervised fine-tuning. For example, In-Context RALM \cite{ram2023incontext} employs a zero-shot approach where an off-the-shelf language model is used to re-rank the top-k documents retrieved by a BM25 retriever. This process involves selecting the document that maximizes the likelihood of the generated text, effectively using the LM's semantic understanding to improve document relevance without requiring additional supervised training. The paper also explores training a dedicated re-ranker using self-supervised learning to further enhance the selection of relevant documents, demonstrating that training a re-ranker with domain-specific data can be more effective than zero-shot re-ranking.

\paragraph{Supervised Re-ranking} Supervised re-rankers involve fine-tuning LLMs on specific ranking datasets. This category can be further divided into models like BERT that process query-document pairs to compute relevance scores, models like T5 that treat ranking as a generation task and use generated tokens to determine relevance, and models like RankLLaMA \cite{ma2024finetuning} that employ a prompt-based approach, focusing on the last token's representation for relevance calculation \cite{zhu2023large}. For instance, the re-ranker in Re2G \cite{glass2022reg} is based on a BERT model trained on labeled data (such as MS MARCO) and fine-tuned to improve the relevance ranking of retrieved documents. FiD-Light \cite{hofstätter2023fidlight} employs a supervised approach where the model is fine-tuned on specific datasets to learn how to re-rank passages effectively using source pointers during autoregressive text generation. The model uses a listwise auto-regressive re-ranking mechanism, trained to identify and re-rank relevant passages based on the output generated during the text generation process. GenRT \cite{xu2024listaware} utilizes a combination of an encoder to capture global list-level features and a sequential decoder to reorder documents based on relevance. The model is trained to learn relevance scores through supervised learning, guided by labeled relevance data, ensuring that the most pertinent documents are prioritized in the final reranked list. Furthermore, ITER-RETGEN \cite{shao2023enhancing} proposes using a more capable re-ranker, which has access to model generations, to distill knowledge into a dense retriever. This knowledge distillation process optimizes the query encoder of the dense retriever, enabling it to better capture the semantic relevance of documents relative to the task input.

\paragraph{Data Augmentation for Re-ranking} Data augmentation for re-rankers focuses on enhancing the training process by generating additional training data, such as pseudo-relevance labels, using LLMs. This data augmentation provides more varied training examples, which helps improve the performance of re-ranking models. For example, DKS-RAC \cite{huang2023retrieval} introduces methods like Dense Knowledge Similarity (DKS) and Retriever as Answer Classifier (RAC), which focus on improving the retrieval process by incorporating rich answer encodings. These methods involve generating additional training signals or utilizing enriched data representations to improve the retrieval and ranking of documents. Additionally, the PROMPTAGATOR \cite{dai2023promptagator} framework utilizes synthetic data generated through LLM-based query generation to enhance the training of the reranker. This data augmentation approach allows the re-ranker to refine candidate passages more effectively, using a cross-attention model trained on these additional examples to boost retrieval accuracy.

\subsection{Filtering}
Filtering and re-ranking are distinct processes in the post-retrieval stage of RAG systems. Filtering focuses on eliminating irrelevant or low-quality documents from the retrieved set, thereby reducing the document set size and improving efficiency and effectiveness in subsequent processing. In contrast, re-ranking orders the remaining documents based on their relevance or utility for the task, often prioritizing those that enhance the quality of the generated output, especially in response-aware scenarios.

Several filtering methods have been developed to refine document sets in RAG systems, each with unique mechanisms but sharing common goals of improving relevance and reducing computational load. Self-RAG \cite{asai2024selfrag} employs a self-reflection mechanism, utilizing special ``reflection tokens'' generated by the model to evaluate the relevance and quality of retrieved passages and the model's own generated outputs. This self-reflection ensures that only the most pertinent documents are retained, leveraging the model's internal capabilities without relying on external models during inference. Similarly, BlendFilter \cite{wang2024blendfilter} utilizes the LLM itself as the filter, assessing and removing irrelevant or less useful documents by applying filtering separately to knowledge retrieved from original, externally augmented, and internally augmented queries. Both Self-RAG and BlendFilter highlight the model's intrinsic ability to perform filtering, reducing the need for additional models and enhancing computational efficiency.

In contrast, RECOMP \cite{xu2024recomp} and CRAG \cite{yan2024corrective} employ more external or structural strategies. RECOMP focuses on selective augmentation, where summaries generated from retrieved documents are selectively prepended to the input for the language model. If the retrieved documents are deemed irrelevant, the compressor can generate an empty summary, effectively filtering out unnecessary information. This method allows for a dynamic approach to filtering, where only helpful content is retained. CRAG, on the other hand, uses a decompose-then-recompose approach, where retrieved documents are split into finer knowledge strips. These strips are evaluated for relevance using a fine-tuned T5 model, and only the relevant strips are recomposed to form a refined set of information for the generation task. This granular filtering process ensures that the final document set is both relevant and concise, tailored specifically to the generation task.

Dynamic filtering techniques are also employed in methods like FiD-TF \cite{berchansky2023optimizing} and CoK \cite{li2024chainofknowledge}. FiD-TF introduces Token Filtering during the decoding process, where less relevant tokens are dynamically filtered out based on cross-attention scores. This approach reduces the computational load by eliminating tokens deemed uninformative for generating the final answer, enhancing efficiency with minimal impact on performance. CoK employs a filtering technique based on self-consistency, identifying and processing only those questions with ``uncertain'' answers. This method works by sampling various reasoning paths and answers, preserving only predictions with high consistency. Questions that do not meet the specified consistency threshold undergo further processing, effectively preventing the propagation of errors in the generation process.

Finally, FILCO \cite{wang2023learning} implements a comprehensive filtering approach using three distinct strategies: String Inclusion (STRINC) to match exact outputs, Lexical Overlap to measure word-level similarity, and Conditional Cross-Mutual Information (CXMI) to assess how much the context improves output likelihood. FILCO applies these filtering strategies at the sentence level, refining the retrieved content for better relevance. Additionally, FILCO trains a context filtering model using these strategies, which predicts the most useful context at inference time, thereby enhancing the accuracy and relevance of the generation model's output.

\section{Generation}
\label{sec:generation}
\subsection{Enhancing}
Enhancing methods are strategies aimed at improving the quality and relevance of generated outputs by integrating retrieved content in various ways. These methods differ in how they combine, aggregate, or refine retrieved information, offering multiple approaches to enrich the final output. Broadly, these techniques can be grouped into three categories: enhancing with queries, enhancing with ensemble approaches, and enhancing with feedback loops.

\paragraph{Enhance with Query}
This approach integrates the retrieved documents with the original query, enabling the generator to leverage both sources in producing the final output. By combining the query with the retrieved content, the generation process ensures that the response remains closely aligned with the user’s intent while being enriched by relevant information. The focus here is on the seamless fusion of the query and context, allowing the generated output to maintain both relevance and completeness. For instance, the RETRO \cite{borgeaud2022improving} model enhances generation by integrating retrieved text chunks with the user’s query using a chunked cross-attention mechanism, where relevant information from the retrieved neighbors is directly injected into the generation process. This method involves first retrieving similar document chunks based on the query and then using a cross-attention module to align and combine these chunks with the input sequence during generation. In-Context RALM \cite{ram2023incontext} takes a comparable approach, directly prepending the retrieved documents to the input query. In this way, the language model can generate responses conditioned on both the query and the retrieved content without requiring changes to the model's architecture. Both examples illustrate a straightforward yet effective method: concatenating the query and retrieved documents into a single input sequence that the LLMs process together, yielding outputs that are contextually enhanced.

\paragraph{Enhance with Ensemble}
When multiple sources are synthesized, the generation process can achieve a more coherent and well-rounded response. Rather than relying solely on a single source, this approach aggregates information from various documents, allowing the generator to reconcile conflicting details, blend diverse perspectives, and select the most reliable or comprehensive output. The ensemble process can manifest in different ways: it may involve combining insights from several sources into a unified narrative, or generating multiple candidate outputs and choosing the best one based on criteria like consistency, relevance, or factual accuracy. An instance of this strategy is seen in FiD \cite{izacard2021leveraging}, which encodes multiple retrieved passages independently before fusing them in the decoder to create a coherent answer. By treating each passage separately during encoding and then merging them during decoding, the model effectively combines evidence from multiple sources. Meanwhile, in REPLUG \cite{shi2023replug}, an ensemble approach is adopted where each retrieved document is independently prepended to the query and processed separately. The outputs are then aggregated, with relevance scores guiding the weighting of each document’s contribution. Through this process, the model capitalizes on diverse information across several sources, leading to improvements in answer accuracy, coverage, and scalability as more data becomes available.

\paragraph{Enhance with Feedback}
In contrast to approaches that process retrieved information in a single pass, this method introduces iterative refinement into the generation process by incorporating feedback loops. Initially, the generator produces a draft response, which is then evaluated and adjusted based on feedback mechanisms, such as self-reflection or predefined criteria focused on factual accuracy and fluency. This iterative approach aims to incrementally improve the output by identifying and correcting errors or fine-tuning content to better align with quality standards, ultimately producing a polished and reliable response. PRCA \cite{yang2023prca} offers an example by positioning itself between the retriever and generator, distilling retrieved information based on feedback from the generator. This distilled information serves as a reward model to guide context optimization, leveraging reinforcement learning and metrics like ROUGE-L scores to iteratively refine which details should be emphasized or downplayed. DSP \cite{khattab2022demonstratesearchpredict}, on the other hand, refines both queries and retrieved passages through a multi-hop retrieval process that incorporates programmatically bootstrapped feedback. Here, the language model generates intermediate queries, retrieves relevant passages, and updates the context in subsequent steps—each stage building on the last to refine the final output. Feedback-driven enhancements are also evident in models like Selfmem \cite{cheng2023lift}, which focus on generating self-memory. The model first produces an unbounded pool of outputs and then selects the most relevant one as memory for the next generation, guided by metrics like BLEU or ROUGE. Finally, RECITE \cite{shi2023replug} integrates feedback by generating multiple recitations from the model’s internal knowledge and using self-consistency techniques to aggregate the outputs. By introducing diversity in the recitations and leveraging passage hints during generation, this approach selects the best content through majority voting. Together, these methods demonstrate how feedback loops and iterative refinements can lead to outputs that are not only more accurate but also increasingly coherent and contextually grounded as they evolve.

\subsection{Customization}
Customization focuses on tailoring content to the user's personality and needs. It involves adjusting the output either to align with specific knowledge retrieved during earlier stages (content alignment) or to adapt the generated response to meet the user’s preferences, context, or audience needs (contextual adaptation).

In LAPDOG \cite{huang2023learning}, customization is achieved primarily through content alignment by integrating persona profiles with external stories to enrich the context used for generation. The story retriever identifies relevant narratives based on the persona, expanding the limited profiles with additional information. The generator then combines this enriched knowledge with the dialogue history, ensuring that responses align closely with the persona’s traits and background. This approach allows for a nuanced understanding of the user’s personality, making the output more engaging and contextually appropriate.

On the other hand, PersonaRAG \cite{zerhoudi2024personarag} emphasizes real-time adaptation by customizing generated content based on dynamic user profiles, session behavior, and ongoing feedback. A multi-agent system continuously analyzes user interactions to refine responses, ensuring alignment with the user’s preferences and context. By integrating personalized insights at each step, the system can adjust its output to suit specific informational needs and situational contexts. This level of responsiveness allows the system to evolve in line with the user’s changing requirements, creating more relevant and targeted responses.

ERAGent \cite{shi2024eragent} also focuses on customization but through the use of a Personalized LLM Reader, which adapts responses using user-specific profiles. This module integrates rewritten questions, filtered knowledge, and user preferences to tailor responses according to both content relevance and user needs. For instance, it takes into account preferences like environmental consciousness or dietary restrictions, ensuring that the generated content is not only aligned with retrieved knowledge but also personalized to the user’s particular values and requirements. This deep level of customization ensures that the output is both relevant and personally meaningful, enhancing user engagement.

ROPG \cite{salemi2024optimization} proposes a dynamic pre- and post-generation retriever selection model, enhancing personalization by aligning the retrieval process with both the input context and the user’s preferences. The pre-generation model determines which retrieval strategy—such as recency-based, keyword matching, or semantic retrieval—is most appropriate before generation begins. By tailoring the retrieval process in this way, the model ensures that the documents retrieved from the user profile closely match the current input, thereby aligning the content with relevant user-specific knowledge. Following this, the post-generation model evaluates the outputs generated by different retrieval strategies and selects the most personalized result. This selection is guided by feedback from the generated content, which is then used to adjust future retrievals. By combining content alignment (through pre-generation retrieval) with contextual adaptation (through post-generation evaluation), this approach offers a comprehensive solution for customization within RAG.

\section{Evaluation in RAG}
\label{sec:evaluation}
\begin{table}
	\centering
	\resizebox{\linewidth}{!}{%
		\begin{tblr}{
				cells = {m},
				cell{2}{1} = {r=3}{},
				cell{2}{2} = {r=3}{},
				cell{2}{5} = {r=3}{},
				cell{5}{1} = {r=3}{},
				cell{5}{2} = {r=3}{},
				cell{5}{4} = {r=3}{},
				cell{5}{5} = {r=3}{},
				cell{8}{1} = {r=2}{},
				cell{8}{2} = {r=2}{},
				cell{8}{5} = {r=2}{},
				cell{10}{1} = {r=4}{},
				cell{10}{2} = {r=4}{},
				cell{10}{5} = {r=4}{},
				cell{14}{1} = {r=4}{},
				cell{14}{2} = {r=4}{},
				cell{14}{4} = {r=4}{},
				cell{14}{5} = {r=4}{},
				cell{18}{1} = {r=3}{},
				cell{18}{2} = {r=3}{},
				cell{18}{5} = {r=3}{},
				cell{21}{1} = {r=2}{},
				cell{21}{2} = {r=2}{},
				cell{21}{5} = {r=2}{},
				vlines,
				hline{1-2,5,8,10,14,18,21,23} = {-}{},
				hline{3-4,9,11-13,19-20,22} = {3-4}{},
				hline{6-7,15-17} = {3}{},
			}
			\textbf{Evaluation Framework} & \textbf{Aspects} & \textbf{Methods} & \textbf{Metrics} & \textbf{Datasets}\\
			RAGAS \cite{es2023ragas} & Quality of RAG Systems & Context Relevance & Extracted Sentences / Total Sentences & WikiEval\footnote{https://huggingface.co/datasets/explodinggradients/WikiEval}\\
			&  & Answer Relevance & Average Cosine Similarity & \\
			&  & Faithfulness & Supported Statements / Total Statements & \\
			ARES \cite{Saad2023ARES} & Improving RAGAS & Context Relevance & Confidence Intervals & {KILT \cite{petroni2021kilt} \\SuperGLUE \cite{wang2019superglue} }\\
			&  & Answer Relevance &  & \\
			&  & Answer Faithfulness &  & \\
			RECALL  \cite{liu2023recall} & Counterfactual Robustness & Response Quality & {Accuracy (QA)\\BLEU, ROUGE-L (Generation)} & {EventKG \cite{gottschalk2018eventkg} \\UJ \cite{huang2022understanding} }\\
			&  & Robustness & {Misleading Rate (QA)\\Mistake Reappearance Rate (Generation)} & \\
			RGB \cite{chen2024benchmarking}& Impact of RAG on LLMs & Noise Robustness & Accuracy & Synthetic\\
			&  & Negative Rejection & Rejection Rate & \\
			&  & Information Integration & Accuracy & \\
			&  & Counterfactual Robustness & {Error Detection Rate \\Error Correction Rate} & \\
			MIRAGE \cite{xiong2024benchmarking}& RAG in Medical QA & Zero-Shot Learning & Accuracy & {MMLU-Med \cite{hendrycks2021measuring}\\MedQA-US \cite{jin2021what}\\MedMCQA \cite{pal2022medmcqa}\\PubMedQA \cite{jin2019pubmedqa}\\BioASQ-Y/N \cite{tsatsaronis2015overview}}\\
			&  & Multi-Choice Evaluation &  & \\
			&  & Retrieval-Augmented Generation &  & \\
			&  & Question-Only Retrieval &  &  \\
			eRAG \cite{salemi2024evaluating} & Retrieval Quality in RAG & Downstream Task & Accuracy, ROUGE & KILT\\
			&  & Set-based & Precision, Recall, Hit Rate & \\
			&  & Ranking & MAP, MRR, NDCG & \\
			BERGEN \cite{rau2024bergen} & Standardizing RAG Experiments & Surface-Based & EM, F1, Precision, Recall & QA Datasets \cite{kwiatkowski2019natural, joshi2017triviaqa}\\
			&  & Semantic & BEM \cite{bulian2022tomayto}, LLMeval \cite{rau2024bergen}& 
		\end{tblr}
	}
	\caption{The Comparison of Different RAG Evaluation Frameworks.}
	\label{tab:evaluation}
\end{table}

To assess how effectively language models can generate more accurate, relevant, and robust responses by leveraging external knowledge, the evaluation of RAG systems has emerged as a crucial research focus. Given the rising popularity of dialogue-based interactions, much recent work has concentrated on evaluating RAG models' performance on such downstream tasks using established metrics like Exact Match (EM) and F1 scores. These metrics have been applied across a wide array of datasets, including TriviaQA \cite{joshi2017triviaqa}, HotpotQA \cite{yang2018hotpotqa}, FEVER \cite{thorne2018fever}, Natural Questions (NQ) \cite{kwiatkowski2019natural}, Wizard of Wikipedia (WoW) \cite{dinan2019wizard}, and T-REX \cite{elsahar2018trex}, which are often used to benchmark the effectiveness of retrieval and generation components in knowledge-intensive tasks.

While downstream task evaluations provide valuable insights, they fail to address the multifaceted challenges that arise as RAG systems continue to evolve. To fill this gap, recent research has proposed various frameworks and benchmarks that aim to evaluate these systems from multiple perspectives, considering not only the quality of the generated text but also the relevance of retrieved documents and the system’s resilience to misinformation, as shown in Table \ref{tab:evaluation}. These evaluations include metrics that assess noise robustness, negative prompting, information integration, and counterfactual robustness, all of which reflect the complex challenges RAG systems face in real-world applications. The ongoing development of comprehensive evaluation frameworks and metrics is essential for advancing the field, broadening the applicability of RAG systems, and ensuring that they meet the demands of an increasingly dynamic and complex information landscape \cite{yu2024evaluation}.

\subsection{Retrieval-based Aspect}
In information retrieval, standard metrics such as Mean Average Precision (MAP), Precision, Reciprocal Rank, and Normalized Discounted Cumulative Gain (NDCG) \cite{radlinski2010comparing, reimers2019sentencebert, nogueira2019multistage} have traditionally been used to evaluate the relevance of retrieved documents to a given query. These metrics are essential in assessing the effectiveness of traditional information retrieval systems, where the primary goal is to measure how well the retrieved documents match the user’s query.

When applied to RAG systems, these retrieval-based metrics extend their focus to consider how the retrieved information contributes to the quality of the generated output. In this context, Accuracy becomes a crucial metric, assessing how precisely the retrieved documents provide correct information for answering queries. Additionally, Rejection Rate \cite{chen2024benchmarking}, which measures the system’s ability to decline answering when no relevant information is available, has emerged as a key indicator of responsible output generation. Similarly, Error Detection Rate \cite{chen2024benchmarking} evaluates the model’s capability to identify and filter out incorrect or misleading information, ensuring that the generation process is based on trustworthy sources.

Another important consideration is Context Relevance, which assesses the alignment of retrieved documents with the specific query, emphasizing the need for content directly relevant to the generation task’s context. Faithfulness \cite{es2023ragas} is also critical in determining whether the generated text accurately reflects the information found in the retrieved documents, thereby minimizing the risk of generating misleading or incorrect content.

The eRAG framework \cite{salemi2024evaluating} introduces a more refined approach to evaluating retrieval quality in RAG systems by focusing on individual documents rather than the entire retrieval process. It operates by feeding each document in the retrieval list into the LLM alongside the query and evaluating the generated output against downstream task metrics such as Accuracy. The document-level scores are then aggregated using ranking metrics like MAP to produce a single evaluation score. This focus on document-level contributions offers a more precise assessment of retrieval quality while being significantly more computationally efficient than traditional end-to-end evaluations.

Notably, eRAG demonstrates that its document-level evaluation correlates more strongly with downstream RAG performance compared to conventional methods like human annotations or provenance labels. This correlation underscores that the LLM, as the primary consumer of the retrieved results, is the most reliable judge of retrieval performance \cite{salemi2024evaluating}. Regardless of the retrieval model or the number of retrieved documents, eRAG consistently outperforms other evaluation approaches, indicating that directly evaluating how each document supports the LLM’s output is the most effective way to measure retrieval quality in RAG systems.

\subsection{Generation-based Aspect}
The evaluation of text produced by large language models involves analyzing performance across a range of downstream tasks using standard metrics that assess linguistic quality, coherence, accuracy, and alignment with ground-truth data. Metrics like BLEU \cite{10.3115/1073083.1073135} and ROUGE-L \cite{lin-2004-rouge} are often used to measure fluency, similarity to human-produced text, and the overlap with reference summaries, respectively, providing insights into how well the generated content captures key ideas and phrases.

In addition to these metrics, which focus on the quality of linguistic output, Accuracy and overlap with ground-truth data are evaluated using EM and F1 scores, which respectively measure the percentage of completely correct answers and offer a balanced view of precision and recall. This ensures that relevant answers are retrieved while inaccuracies are minimized.

Beyond these standard evaluation techniques, more specialized criteria have been introduced to assess RAG systems in specific contexts. For dialogue generation, for instance, metrics like perplexity and entropy are employed to evaluate response diversity and naturalness. In scenarios where misinformation is a concern, metrics like Misleading Rate and Mistake Reappearance Rate \cite{liu2023recall} have been developed to measure a model’s ability to avoid generating incorrect or misleading content. Other advanced metrics include Answer Relevance \cite{es2023ragas}, which assesses the precision of responses to queries, Kendall’s tau \cite{Saad2023ARES}, used for evaluating the accuracy of system rankings, and Micro-F1 \cite{Saad2023ARES}, which fine-tunes accuracy evaluation in tasks involving multiple correct answers. Prediction Accuracy further complements these by directly measuring how closely the generated responses align with the expected answers, offering a clear measure of a system’s effectiveness in producing accurate content.

\begin{table}[htb]
	\centering
	\resizebox{\linewidth}{!}{%
		\begin{tabular}{|c|c|c|c|c|c|c|c|c|c|c|c|c|c|} 
			\hline
			\multirow{2}{*}{\textbf{Research}} & \multirow{2}{*}{\textbf{Year}} & \multicolumn{2}{c|}{\textbf{Retrieval Source}} & \multirow{2}{*}{\textbf{Multi-hop}} & \multirow{2}{*}{\textbf{Training}} & \multicolumn{3}{c|}{\textbf{Pre-Retrieval}} & \textbf{Retrieval} & \multicolumn{2}{c|}{\textbf{Post-Retrieval}} & \multicolumn{2}{c|}{\textbf{Generation}} \\ 
			\cline{3-4}\cline{7-14}
			&  & \textbf{Internal} & \textbf{External} &  &  & \textbf{Indexing} & \textbf{Query Manipulation} & \textbf{Data Modification} & \textbf{Search \& Ranking} & \textbf{Re-Ranking} & \textbf{Filtering} & \textbf{Enhancing} & \textbf{Customization} \\ 
			\hline
			REALM \cite{guu2020retrieval} & 2020 &  & \Checkmark &  & \Checkmark & \Checkmark &  &  & \Checkmark &  &  &  &  \\ 
			\hline
			kNN-LMs \cite{khandelwal2020generalization} & 2020 & \Checkmark & \Checkmark &  &  & \Checkmark &  &  & \Checkmark &  &  & \Checkmark &  \\ 
			\hline
			RAG \cite{lewis2020retrievalaugmented} & 2020 &  & \Checkmark &  & \Checkmark & \Checkmark &  &  & \Checkmark &  &  &  &  \\ 
			\hline
			FiD \cite{izacard2021leveraging} & 2021 &  & \Checkmark &  &  &  &  &  & \Checkmark &  &  & \Checkmark &  \\ 
			\hline
			Webgpt \cite{nakano2021webgpt} & 2021 &  & \Checkmark & \Checkmark & \Checkmark & \Checkmark & \Checkmark &  & \Checkmark &  & \Checkmark & \Checkmark &  \\ 
			\hline
			Re2G \cite{glass2022reg} & 2022 & \Checkmark &  & \Checkmark & \Checkmark &  &  &  &  & \Checkmark &  &  &  \\ 
			\hline
			RETRO \cite{borgeaud2022improving} & 2022 &  & \Checkmark & \Checkmark & \Checkmark & \Checkmark &  &  & \Checkmark &  &  &  &  \\ 
			\hline
			DSP \cite{khattab2022demonstratesearchpredict} & 2022 &  & \Checkmark & \Checkmark &  &  & \Checkmark &  &  & \Checkmark &  & \Checkmark &  \\ 
			\hline
			CoK \cite{li2024chainofknowledge} & 2023 &  & \Checkmark & \Checkmark &  &  & \Checkmark &  &  & \Checkmark &  &  &  \\ 
			\hline
			IRCOT \cite{trivedi2023interleaving} & 2023 &  & \Checkmark & \Checkmark &  &  & \Checkmark &  &  &  &  & \Checkmark &  \\ 
			\hline
			ITRG \cite{feng2024retrievalgeneration} & 2023 & \Checkmark & \Checkmark & \Checkmark &  &  &  &  & \Checkmark &  &  & \Checkmark &  \\ 
			\hline
			PKG \cite{luo2023augmented} & 2023 & \Checkmark &  &  &  &  &  &  &  &  &  &  & \\ 
			\hline
			RA-DIT \cite{lin2024radit} & 2023 &  & \Checkmark & \Checkmark & \Checkmark &  &  & \Checkmark & \Checkmark &  &  & \Checkmark &  \\ 
			\hline
			Self-RAG \cite{asai2024selfrag} & 2023 &  & \Checkmark &  & \Checkmark &  &  &  &  &  & \Checkmark &  & \\ 
			\hline
			SURGE \cite{kang2023knowledge} & 2023 & \Checkmark &  &  &  &  &  &  & \Checkmark &  &  &  & \\ 
			\hline
			FiD-TF \cite{berchansky2023optimizing} & 2023 &  & \Checkmark &  &  &  &  &  &  & \Checkmark & \Checkmark &  &  \\ 
			\hline
			PRCA \cite{yang2023prca} & 2023 &  & \Checkmark &  & \Checkmark &  &  &  & \Checkmark &  &  & \Checkmark &  \\ 
			\hline
			REPLUG \cite{shi2023replug} & 2023 &  & \Checkmark &  & \Checkmark &  &  &  &  &  &  & \Checkmark & \\ 
			\hline
			AAR \cite{yu2023augmentationadapted} & 2023 &  & \Checkmark &  & \Checkmark &  &  &  & \Checkmark &  &  &  &  \\ 
			\hline
			Query2doc \cite{wang2023querydoc} & 2023 & \Checkmark &  &  &  &  & \Checkmark &  &  &  &  &  &  \\ 
			\hline
			Step-Back \cite{zheng2024take} & 2023 &  & \Checkmark & \Checkmark &  &  & \Checkmark &  &  &  &  &  &  \\ 
			\hline
			ITER-RETGEN \cite{shao2023enhancing} & 2023 &  & \Checkmark & \Checkmark &  &  &  &  & \Checkmark & \Checkmark &  &  &  \\ 
			\hline
			RECITE \cite{sun2023recitationaugmented} & 2023 & \Checkmark &  & \Checkmark & \Checkmark &  &  & \Checkmark &  &  &  & \Checkmark &  \\ 
			\hline
			PROMPTAGATOR \cite{dai2023promptagator} & 2023 & \Checkmark &  & \Checkmark &  &  & \Checkmark &  &  & \Checkmark & \Checkmark &  &  \\ 
			\hline
			UPRISE \cite{cheng2023uprise} & 2023 & \Checkmark &  & \Checkmark & \Checkmark &  &  & \Checkmark & \Checkmark &  &  & \Checkmark &  \\ 
			\hline
			GENREAD \cite{yu2023generate} & 2023 & \Checkmark &  &  &  &  &  & \Checkmark &  &  &  & \Checkmark &  \\ 
			\hline
			LAPDOG \cite{huang2023learning} & 2023 & & \Checkmark & & \Checkmark & & \Checkmark & & \Checkmark & \Checkmark & & \Checkmark & \Checkmark \\
			\hline
			KnowledGPT \cite{wang2023knowledgpt} & 2023 &  & \Checkmark & \Checkmark &  &  & \Checkmark & \Checkmark &  &  &  &  &  \\ 
			\hline
			Selfmem \cite{cheng2023lift} & 2023 &  & \Checkmark & \Checkmark & \Checkmark &  &  &  &  & \Checkmark &  & \Checkmark &  \\ 
			\hline
			MEMWALKER \cite{chen2023walking} & 2023 &  & \Checkmark &  &  & \Checkmark &  &  & \Checkmark &  &  & \Checkmark &  \\ 
			\hline
			RECOMP \cite{xu2024recomp} & 2023 &  & \Checkmark &  & \Checkmark &  &  &  &  &  & \Checkmark &  &  \\ 
			\hline
			Rewrite-Retrieve-Read \cite{ma2023query} & 2023 &  & \Checkmark &  & \Checkmark &  & \Checkmark &  &  &  &  &  &  \\ 
			\hline
			Atlas \cite{ma2023query} & 2023 &  & \Checkmark & \Checkmark & \Checkmark & \Checkmark &  &  & \Checkmark & \Checkmark &  &  &  \\ 
			\hline
			DKS-RAC \cite{huang2023retrieval} & 2023 &  & \Checkmark & \Checkmark & \Checkmark &  &  &  &  & \Checkmark & \Checkmark &  &  \\ 
			\hline
			In-Context RALM \cite{ram2023incontext} & 2023 &  & \Checkmark &  &  &  &  &  &  & \Checkmark &  &  &  \\ 
			\hline
			Fid-light \cite{hofstätter2023fidlight} & 2023 &  & \Checkmark & \Checkmark &  &  &  &  &  & \Checkmark &  &  &  \\ 
			\hline
			FLARE \cite{jiang2023active} & 2023 &  & \Checkmark &  &  &  & \Checkmark &  & \Checkmark &  &  &  &  \\
			\hline
			Chameleon \cite{jiang2023chameleon} & 2023 & & \Checkmark & & \Checkmark & \Checkmark & & & \Checkmark & & & & \\
			\hline
			ERAGent \cite{shi2024eragent} & 2024 & \Checkmark & \Checkmark &&\Checkmark& &\Checkmark&&\Checkmark&&\Checkmark&\Checkmark&\Checkmark \\
			\hline
			PipeRAG \cite{jiang2024piperag} & 2024 & & \Checkmark & \Checkmark & \Checkmark & \Checkmark & & & & & & \Checkmark & \\
			\hline
			GenRT \cite{xu2024listaware} & 2024 & \Checkmark & & & \Checkmark & & & & & \Checkmark & \Checkmark & & \\
			\hline
			PersonaRAG \cite{zerhoudi2024personarag} & 2024 & \Checkmark & \Checkmark & \Checkmark & \Checkmark & & \Checkmark & & \Checkmark & \Checkmark & & \Checkmark & \Checkmark \\
			\hline
			CRAG \cite{yan2024corrective} & 2024 & \Checkmark & \Checkmark & & \Checkmark & & & \Checkmark & & \Checkmark & \Checkmark & \Checkmark & \\
			\hline
			IMRAG \cite{yang2024imrag} & 2024 & & \Checkmark & \Checkmark & \Checkmark & & \Checkmark & & & \Checkmark & & \Checkmark & \\
			\hline
			AiSAQ \cite{tatsuno2024aisaq} & 2024 & \Checkmark & & & & \Checkmark & & & \Checkmark & \Checkmark & & & \\
			\hline
			ROPG \cite{salemi2024optimization} & 2024 & \Checkmark & & & \Checkmark & & & & \Checkmark & & & \Checkmark & \Checkmark \\
			\hline
			RQ-RAG \cite{chan2024rqrag} & 2024 & & \Checkmark & \Checkmark & \Checkmark & & \Checkmark & & & \Checkmark & & & \\
			\hline
			PlanRAG \cite{lee2024planrag} & 2024 & & \Checkmark & \Checkmark & & & \Checkmark & & & \Checkmark & & \Checkmark & \\
			\hline
			RARG \cite{yue2024evidencedriven} & 2024 & & \Checkmark & & \Checkmark & & & \Checkmark & \Checkmark & & & \Checkmark & \\
			\hline 
			DRAGIN \cite{su2024dragin} & 2024 & & \Checkmark & \Checkmark & & & \Checkmark & & \Checkmark & & & \Checkmark & \\
			\hline
			LRUS-CoverTree \cite{ma2024reconsidering} & 2024 & \Checkmark & & & \Checkmark & \Checkmark & & & \Checkmark & \Checkmark & & & \\
			\hline
		\end{tabular}
	}
	\caption{The comprehensive summary of RAG studies. A \Checkmark in the ``Multi-hop'' column signifies that the research involves multiple search rounds. Similarly, a \Checkmark in the ``Training'' column indicates that the study included training phases. It is important to note that in this context, ``Training'' encompasses both initial model training and fine-tuning processes.}
	\label{tab:appendixb}
\end{table}

\section{Comparisons of RAG}
\label{sec:compare}
\subsection{The Comprehensive Summary of RAG}

Table \ref{tab:appendixb} presents a detailed analysis of the RAG studies discussed in this paper. The analysis shows that the majority of these studies have utilized external data sources to enrich the content of LLMs. A preference for multiple-hop over single-hop retrieval was noted, indicating that iterative search rounds generally yield superior results. In other words, most methods employ dense retrieval to secure higher quality candidate documents. Compared to modifying datasets in the pre-retrieval stage, more studies focus on manipulating the query to improve retrieval performance. Additionally, there is a significant emphasis on optimizing the retrieval phase, highlighting its crucial role in the research. However, there seems to be a scarcity of studies concentrating on customization in the generation stage, pointing to this as a potential area for future exploration. Overall, while the goal of RAG is to enhance the response quality of LLMs, greater efforts have been directed towards improving retrieval aspects.

\begin{table}
	\scriptsize
	\centering
	\resizebox{\linewidth}{!}{%
		\begin{tabular}{|c|c|m{5cm}|m{6cm}|} 
			\hline
			\textbf{Research} & \textbf{Year} & \textbf{Retriever} & \textbf{Generator} \\ 
			\hline
			REALM \cite{guu2020retrieval} & 2020 & BERT \cite{devlin2019bert} & Transformers \cite{vaswani2017attention} \\ 
			\hline
			kNN-LMs \cite{khandelwal2020generalization} & 2020 & FAISS \cite{8733051} & Transformers \\ 
			\hline
			RAG \cite{lewis2020retrievalaugmented} & 2020 & DPR \cite{karpukhin2020dense} & BART-Large \cite{lewis2020bart} \\ 
			\hline
			FiD \cite{izacard2021leveraging} & 2021 & BM25 \cite{robertson2009probabilistic}, DPR & T5 \cite{raffel2020exploring} \\ 
			\hline
			Webgpt \cite{nakano2021webgpt} & 2021 & Bing & GPT-3 \cite{brown2020language} \\ 
			\hline
			Re2G \cite{glass2022reg} & 2022 & BM25, DPR & BART \\ 
			\hline
			RETRO \cite{borgeaud2022improving} & 2022 & BERT & Transformer \\ 
			\hline
			DSP \cite{khattab2022demonstratesearchpredict} & 2022 & ColBERTv2 \cite{khattab2020colbert} & GPT-3.5 (text-davinci-002) \\ 
			\hline
			CoK \cite{li2024chainofknowledge} & 2023 & LLaMA2-7B \cite{touvron2023llama}, ChatGPT (gpt-3.5-turbo-0613) & ChatGPT (gpt-3.5-turbo-0613) \\ 
			\hline
			IRCOT \cite{trivedi2023interleaving} & 2023 & BM25 & GPT-3 (code-davinci-002), Flan-T5~\cite{chung2022scaling} \\ 
			\hline
			ITRG \cite{feng2024retrievalgeneration} & 2023 & Atlas \cite{ma2023query} & LLaMA-33B \\ 
			\hline
			PKG \cite{luo2023augmented} & 2023 & LLaMA-7B & InstructGPT-3.5 (text-davinic-002) \cite{ouyang2022training} \\ 
			\hline
			RA-DIT \cite{lin2024radit} & 2023 & DRAGON+ \cite{lin2023train} & LLaMA \\ 
			\hline
			Self-RAG \cite{asai2024selfrag} & 2023 & Contriever \cite{izacard2022unsupervised} & LLaMA2 (7B and 13B) , GPT-4 \cite{achiam2023gpt} \\ 
			\hline
			SURGE \cite{kang2023knowledge} & 2023 & Graph Neural Networks (GNN) \cite{hamilton2020graph} & Transformers \\ 
			\hline
			FiD-TF \cite{berchansky2023optimizing} & 2023 & BM25, SBERT \cite{reimers2019sentencebert} & T5 \\ 
			\hline
			PRCA \cite{yang2023prca} & 2023 & BM25, DPR, Contriver, SimCSE \cite{gao2021simcse}, SBERT & T5, Phoenix-7B \cite{chen2023phoenix}, Vicuna-7B \cite{vicuna2023}, ChatGLM \cite{du2022glm}, GPT-3.5 \\ 
			\hline
			REPLUG \cite{shi2023replug} & 2023 & Contriever & GPT-3 \\ 
			\hline
			AAR \cite{yu2023augmentationadapted} & 2023 & ANCE \cite{xiong2021approximate}, Contriever & Flan-T5, InstructGPT \\ 
			\hline
			Query2doc \cite{wang2023querydoc} & 2023 & BM25, DPR & GPT-3 (text-davinci-003) \\ 
			\hline
			Step-Back \cite{zheng2024take} & 2023 & PaLM-2L \cite{chowdhery2023palm} & PaLM-2L, GPT-4 \\ 
			\hline
			ITER-RETGEN \cite{shao2023enhancing} & 2023 & Contriever & InstructGPT (text-davinci-003), LLaMA2 \\ 
			\hline
			RECITE \cite{sun2023recitationaugmented} & 2023 & & PaLM, UL2 \cite{tay2023ul}, OPT \cite{zhang2022opt}, Codex \cite{chen2021evaluating} \\ 
			\hline
			PROMPTAGATOR \cite{dai2023promptagator} & 2023 & T5 & FLAN \\ 
			\hline
			UPRISE \cite{cheng2023uprise} & 2023 & GPT-Neo-2.7B \cite{black2022gptneoxb} & BLOOM-7.1B \cite{workshop2022bloom}, OPT-66B, GPT-3-175B \\ 
			\hline
			GENREAD \cite{yu2023generate} & 2023 &  & InstructGPT \\ 
			\hline
			LAPDOG \cite{huang2023learning} & 2023 & Contriever & T5 \\
			\hline
			KnowledGPT \cite{wang2023knowledgpt} & 2023 &  & GPT-4 \\ 
			\hline
			Selfmem \cite{cheng2023lift} & 2023 & BM25 & XGLM \cite{lin2022fewshot}, XLM-Rbase \cite{conneau2020unsupervised} \\ 
			\hline
			MEMWALKER \cite{chen2023walking} & 2023 & LLaMA2 & LLaMA2 \\ 
			\hline
			RECOMP \cite{xu2024recomp} & 2023 & BM25 & T5-Large \\ 
			\hline
			Rewrite-Retrieve-Read \cite{ma2023query} & 2023 & Bing & T5-Large, ChatGPT(gpt-3.5-turbo), Vicuna-13B \\ 
			\hline
			Atlas \cite{ma2023query} & 2023 & Contriever & T5 \\ 
			\hline
			DKS-RAC \cite{huang2023retrieval} & 2023 & DPR & BART \\ 
			\hline
			In-Context RALM \cite{ram2023incontext} & 2023 & BM25, BERT-base, Contriever, Spider \cite{ram2022learning} & GPT-2, GPT-Neo, GPT-J \cite{gao2021pile}, OPT, and LLaMA \\ 
			\hline
			Fid-light \cite{hofstätter2023fidlight} & 2023 & GTR-Base \cite{ni2022large} & T5 \\ 
			\hline
			FLARE \cite{jiang2023active} & 2023 & BM25, Bing & GPT-3.5 (text-davinci-003) \\
			\hline
			Chameleon \cite{jiang2023chameleon} & 2023 & ChamVS \cite{jiang2023chameleon} & ChamLM \cite{jiang2023chameleon} \\
			\hline
			ERAGent \cite{shi2024eragent} & 2024 & Bing & GPT-3.5, Falcon 1B \cite{penedo2023refinedweb} \\
			\hline
			PipeRAG \cite{jiang2024piperag} & 2024 & SBERT & RETRO \cite{borgeaud2022improving} \\
			\hline
			GenRT \cite{xu2024listaware} & 2024 & LambdaMart \cite{ai2018learning} &  \\
			\hline
			PersonaRAG \cite{zerhoudi2024personarag} & 2024 & BM25 & GPT-3.5\\
			\hline
			CRAG \cite{yan2024corrective} & 2024 & Contriever &  LLaMA2\\
			\hline
			IMRAG \cite{yang2024imrag} & 2024 & DPR & Vicuna-7B  \\
			\hline
			AiSAQ \cite{tatsuno2024aisaq} & 2024 & DiskANN \cite{pan2023lmdiskann} &  \\
			\hline
			ROPG \cite{salemi2024optimization} & 2024 & BM25, Contriever & FlanT5-XXL \\
			\hline
			RQ-RAG \cite{chan2024rqrag} & 2024 & DuckDuckGo\footnote{https://duckduckgo.com} & LLaMA2-7B \\
			\hline
			PlanRAG \cite{lee2024planrag} & 2024 & GPT-4 & GPT-4 \\
			\hline
			RARG \cite{yue2024evidencedriven} & 2024 & BM25, E5 \cite{wang2022text}& LLaMA2-7B  \\
			\hline 
			DRAGIN \cite{su2024dragin} & 2024 & BM25, SGPT \cite{muennighoff2022sgpt} & LLaMA2 (7B and 13B), Vicuna-13B  \\
			\hline
			LRUS-CoverTree \cite{ma2024reconsidering} & 2024 & k-MIPS & \\
			\hline
		\end{tabular}
	}
	\caption{The summary of Retrievers and Generators. The retrieval models and pre-trained language models explicitly mentioned in these studies have been recorded.}
	\label{tab:regencomp}
\end{table}

\begin{table}
	\centering
	\resizebox{\linewidth}{!}{
	\begin{tabular}{cccccccccc}
		\toprule
		\multirow{3}{*}{\bf Corpus} & \multirow{3}{*}{\bf Retriever} & \multicolumn{5}{c}{\bf \textsc{Mirage} Benchmark Dataset} & \multirow{3}{*}{\bf Average} \\ 
		\cmidrule(lr){3-7}
		&  & \bf MMLU-Med & \bf MedQA-US & \bf MedMCQA & \bf PubMedQA* & \bf BioASQ-Y/N &  \\
		\midrule
		None & None & 72.91 \textcolor{gray}{\scriptsize $\pm$ 1.35} & 65.04 \textcolor{gray}{\scriptsize $\pm$ 1.34} & 55.25 \textcolor{gray}{\scriptsize $\pm$ 0.77} & 36.00 \textcolor{gray}{\scriptsize $\pm$ 2.15} & 74.27 \textcolor{gray}{\scriptsize $\pm$ 1.76} & 60.69 \\
		\midrule
		\multirow{6}{*}{\makecell{\textbf{PubMed}\\ (23.9M)}} 
		& BM25 & \cellcolor{red!40.4!} 72.27 \textcolor{gray}{\scriptsize $\pm$ 1.36} & \cellcolor{red!67.5!} 63.71 \textcolor{gray}{\scriptsize $\pm$ 1.35} & \cellcolor{green!6.8!} 55.49 \textcolor{gray}{\scriptsize $\pm$ 0.77} & \cellcolor{green!73.2!} 66.20 \textcolor{gray}{\scriptsize $\pm$ 2.12} & \cellcolor{green!71.1!} 88.51 \textcolor{gray}{\scriptsize $\pm$ 1.28} & \cellcolor{green!62.8!} 69.23 \\
		& Contriever & \cellcolor{red!74.1!} 71.72 \textcolor{gray}{\scriptsize $\pm$ 1.36} & \cellcolor{red!55.5!} 63.94 \textcolor{gray}{\scriptsize $\pm$ 1.35} & \cellcolor{red!30.1!} 54.29 \textcolor{gray}{\scriptsize $\pm$ 0.77} & \cellcolor{green!71.7!} 65.60 \textcolor{gray}{\scriptsize $\pm$ 2.12} & \cellcolor{green!55.7!} 85.44 \textcolor{gray}{\scriptsize $\pm$ 1.42} & \cellcolor{green!55.2!} 68.20 \\
		& SPECTER & \cellcolor{green!5.5!} 73.19 \textcolor{gray}{\scriptsize $\pm$ 1.34} & \cellcolor{green!5.2!} 65.20 \textcolor{gray}{\scriptsize $\pm$ 1.34} & \cellcolor{red!66.8!} 53.12 \textcolor{gray}{\scriptsize $\pm$ 0.77} & \cellcolor{green!45.6!} 54.80 \textcolor{gray}{\scriptsize $\pm$ 2.23} & \cellcolor{green!7.3!} 75.73 \textcolor{gray}{\scriptsize $\pm$ 1.72} & \cellcolor{green!27.3!} 64.41 \\
		& MedCPT & \cellcolor{green!3.5!} 73.09 \textcolor{gray}{\scriptsize $\pm$ 1.34} & \cellcolor{green!54.0!} 66.69 \textcolor{gray}{\scriptsize $\pm$ 1.32} & \cellcolor{red!9.8!} 54.94 \textcolor{gray}{\scriptsize $\pm$ 0.77} & \cellcolor{green!73.7!} 66.40 \textcolor{gray}{\scriptsize $\pm$ 2.11} & \cellcolor{green!57.3!} 85.76 \textcolor{gray}{\scriptsize $\pm$ 1.41} & \cellcolor{green!63.8!} 69.38 \\
		& RRF-2 & \cellcolor{green!56.3!} 75.57 \textcolor{gray}{\scriptsize $\pm$ 1.30} & \cellcolor{red!35.6!} 64.34 \textcolor{gray}{\scriptsize $\pm$ 1.34} & \cellcolor{green!2.7!} 55.34 \textcolor{gray}{\scriptsize $\pm$ 0.77} & \cellcolor{green!80.0!} 69.00 \textcolor{gray}{\scriptsize $\pm$ 2.07} & \cellcolor{green!63.8!} 87.06 \textcolor{gray}{\scriptsize $\pm$ 1.35} & \cellcolor{green!70.3!} 70.26 \\
		& RRF-4 & \cellcolor{green!9.4!} 73.37 \textcolor{gray}{\scriptsize $\pm$ 1.34} & \cellcolor{red!15.7!} 64.73 \textcolor{gray}{\scriptsize $\pm$ 1.34} & \cellcolor{red!15.8!} 54.75 \textcolor{gray}{\scriptsize $\pm$ 0.77} & \cellcolor{green!75.6!} 67.20 \textcolor{gray}{\scriptsize $\pm$ 2.10} & \cellcolor{green!71.1!} 88.51 \textcolor{gray}{\scriptsize $\pm$ 1.28} & \cellcolor{green!66.3!} 69.71 \\
		\midrule
		
		\multirow{6}{*}{\makecell{\textbf{Wikipedia}\\(29.9M)}} 
		& BM25 & \cellcolor{green!9.4!} 73.37 \textcolor{gray}{\scriptsize $\pm$ 1.34} & \cellcolor{red!79.4!} 63.47 \textcolor{gray}{\scriptsize $\pm$ 1.35} & \cellcolor{red!36.1!} 54.10 \textcolor{gray}{\scriptsize $\pm$ 0.77} & \cellcolor{red!55.6!} 26.40 \textcolor{gray}{\scriptsize $\pm$ 1.97} & \cellcolor{red!12.4!} 71.36 \textcolor{gray}{\scriptsize $\pm$ 1.82} & \cellcolor{red!37.8!} 57.74 \\
		& Contriever & \cellcolor{green!25.0!} 74.10 \textcolor{gray}{\scriptsize $\pm$ 1.33} & \cellcolor{green!30.9!} 65.99 \textcolor{gray}{\scriptsize $\pm$ 1.33} & \cellcolor{red!38.3!} 54.03 \textcolor{gray}{\scriptsize $\pm$ 0.77} & \cellcolor{red!55.6!} 26.40 \textcolor{gray}{\scriptsize $\pm$ 1.97} & \cellcolor{red!18.6!} 69.90 \textcolor{gray}{\scriptsize $\pm$ 1.85} & \cellcolor{red!33.4!} 58.08 \\
		& SPECTER & \cellcolor{red!46.0!} 72.18 \textcolor{gray}{\scriptsize $\pm$ 1.36} & \cellcolor{red!71.4!} 63.63 \textcolor{gray}{\scriptsize $\pm$ 1.35} & \cellcolor{red!79.6!} 52.71 \textcolor{gray}{\scriptsize $\pm$ 0.77} & \cellcolor{red!79.9!} 22.20 \textcolor{gray}{\scriptsize $\pm$ 1.86} & \cellcolor{red!31.7!} 66.83 \textcolor{gray}{\scriptsize $\pm$ 1.89} & \cellcolor{red!66.3!} 55.51 \\
		& MedCPT & \cellcolor{red!57.2!} 71.99 \textcolor{gray}{\scriptsize $\pm$ 1.36} & \cellcolor{green!2.7!} 65.12 \textcolor{gray}{\scriptsize $\pm$ 1.34} & \cellcolor{red!3.1!} 55.15 \textcolor{gray}{\scriptsize $\pm$ 0.77} & \cellcolor{red!40.6!} 29.00 \textcolor{gray}{\scriptsize $\pm$ 2.03} & \cellcolor{red!3.4!} 73.46 \textcolor{gray}{\scriptsize $\pm$ 1.78} & \cellcolor{red!22.3!} 58.95 \\
		& RRF-2 & \cellcolor{green!26.9!} 74.20 \textcolor{gray}{\scriptsize $\pm$ 1.33} & \cellcolor{red!23.7!} 64.57 \textcolor{gray}{\scriptsize $\pm$ 1.34} & \cellcolor{red!16.6!} 54.72 \textcolor{gray}{\scriptsize $\pm$ 0.77} & \cellcolor{red!29.0!} 31.00 \textcolor{gray}{\scriptsize $\pm$ 2.07} & \cellcolor{green!9.7!} 76.21 \textcolor{gray}{\scriptsize $\pm$ 1.71} & \cellcolor{red!7.0!} 60.14 \\
		& RRF-4 & \cellcolor{green!5.5!} 73.19 \textcolor{gray}{\scriptsize $\pm$ 1.34} & \cellcolor{red!3.8!} 64.96 \textcolor{gray}{\scriptsize $\pm$ 1.34} & \cellcolor{red!22.6!} 54.53 \textcolor{gray}{\scriptsize $\pm$ 0.77} & \cellcolor{red!29.0!} 31.00 \textcolor{gray}{\scriptsize $\pm$ 2.07} & \cellcolor{red!9.6!} 72.01 \textcolor{gray}{\scriptsize $\pm$ 1.81} & \cellcolor{red!19.9!} 59.14 \\
		
		\bottomrule
	\end{tabular}
	}
	\caption{Part results of Accuracy (\%) of GPT-3.5 across different corpora and retrievers on Mirage. Red and green highlight \colorbox{red!50}{declines} and \colorbox{green!50}{improvements} compared to CoT (first row), with shading intensity reflecting the degree of change. Data sourced from Mirage \cite{xiong2024benchmarking}.}
	\label{tab:retrievalcomp}
\end{table}

\begin{figure}
	\centering
	\begin{subfigure}[b]{0.45\linewidth}
		\centering
		\includegraphics[width=\linewidth]{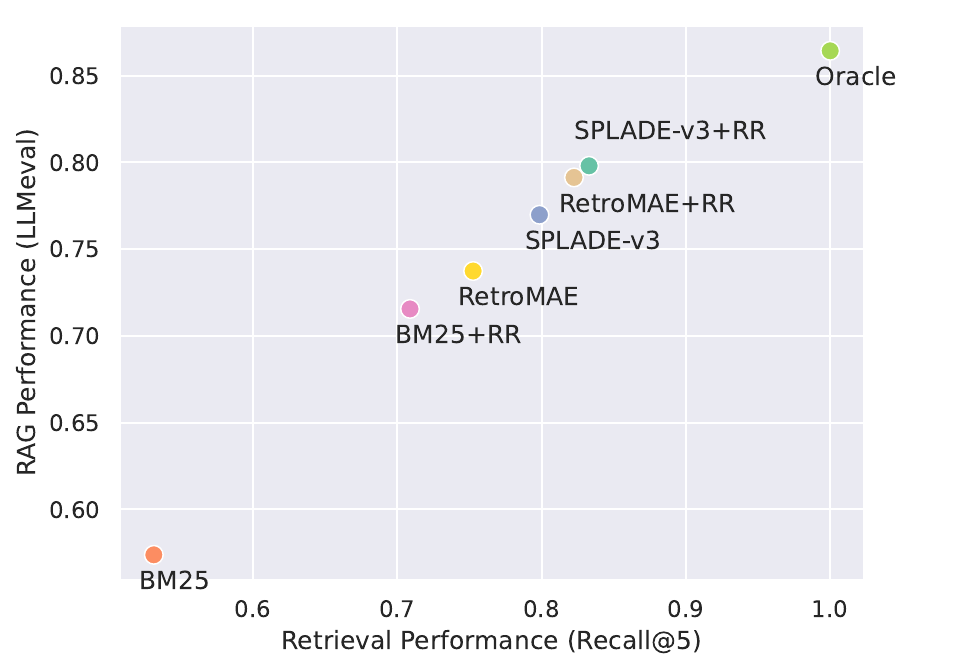}
		\caption{Impact of retrieval performance on RAG performance for SOLAR-10.7B \cite{kim2024solar} on NQ with different ranking systems. RR means with additional re-ranking using DeBERTa-v3.}
		\label{fig:regencomp01}
	\end{subfigure}
	\hfill
	\begin{subfigure}[b]{0.45\linewidth}
		\centering
		\includegraphics[width=\linewidth]{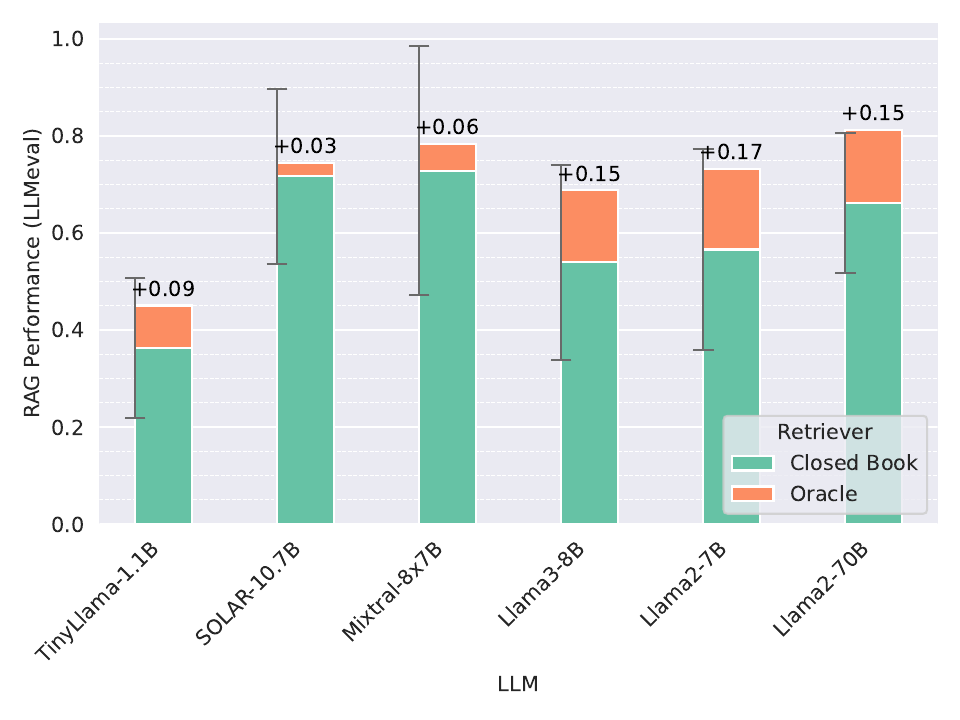}
		\caption{Performance gains w/ and w/o oracle retrieval for LLMs with different sizes. Comparing closed book vs oracle passages averaged over all QA datasets in KILT.}
		\label{fig:regencomp02}
	\end{subfigure}
	
	\begin{subfigure}[b]{\linewidth}
		\centering
		\includegraphics[width=0.9\linewidth]{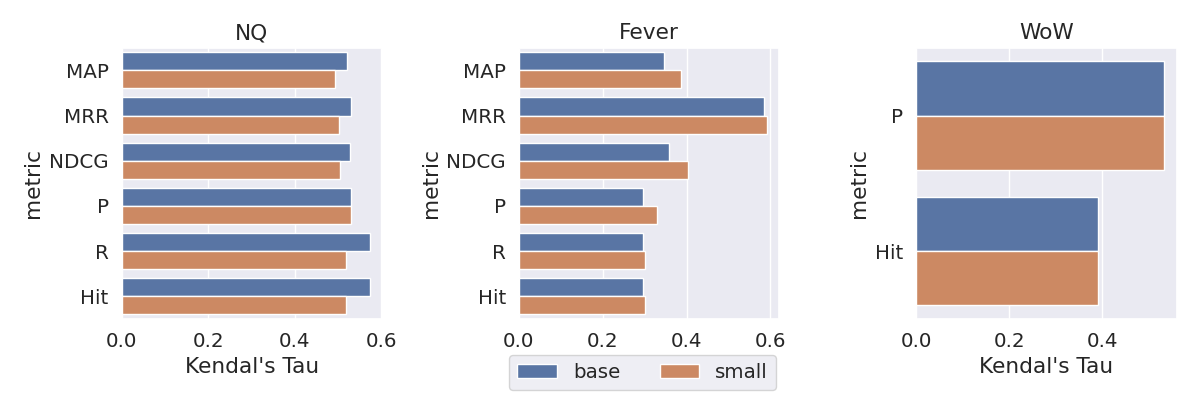}
		\caption{The correlation between eRAG and the downstream performance of different LLM sizes. In this experiment, T5small (60M parameters) and T5-base (220M parameters) with FiD are used. The documents are retrieved using BM25.}
		\label{fig:regencomp03}
	\end{subfigure}
	
	\caption{Retriever and generator experiment results sourced from eRAG \cite{salemi2024evaluating} and BERGEN \cite{rau2024bergen}.}
	\label{fig:regencomp}
\end{figure}

\subsection{Retriever and Generator}
In RAG, the retriever and generator are central components, each playing a distinct role in the system's overall performance. Table \ref{tab:regencomp} summarizes the retrievers and generators used across the studies discussed in this paper. The table reveals that while a wide range of advanced language models are employed as generators, many systems still rely on traditional retrievers like BM25, valued for their efficiency. This highlights the continued importance of optimizing retrieval methods while balancing computational demands. Interestingly, despite the availability of powerful models such as LLaMA2, GPT-3.5, and GPT-4, these are not widely adopted as generators. Instead, models like T5 remain prevalent, while more foundational retrieval approaches, such as those based on BERT, see limited use. The relative scarcity of IR-focused LLMs in retrievers suggests a promising avenue for future research and development in this domain.

\paragraph{Impact of the Retriever} The results shown in Table \ref{tab:retrievalcomp} highlight the accuracy of GPT-3.5 across different corpora and retrievers on the Mirage benchmark \cite{xiong2024benchmarking}. These findings underscore how retriever performance closely depends on the alignment between training data and the target corpus. For example, in the MEDRAG system, MedCPT—trained specifically on PubMed user logs—significantly improves retrieval performance when accessing the PubMed corpus. This illustrates the benefits of using domain-specific retrievers tailored to specialized datasets. In contrast, general-purpose retrievers like Contriever, which incorporate Wikipedia data during training, excel in retrieving information from Wikipedia, especially for tasks like MMLU-Med and MedQA-US. On the other hand, SPECTER, which focuses more on regularizing pairwise article distances than optimizing query-to-article relevance, underperforms on the MedCorp corpus. The study also explores combining multiple retrievers using Reciprocal Rank Fusion (RRF). However, results show that adding more retrievers does not always lead to better outcomes; for instance, excluding SPECTER in RRF-2 on Wikipedia yields better results than RRF-4, indicating that simply increasing the number of retrievers is not beneficial unless their strengths align with the retrieval task.

Figure \ref{fig:regencomp01} illustrates how eRAG investigates the correlation between LLM performance and retrieval effectiveness on the NQ dataset using three retrievers with different characteristics: BM25 (lexical sparse), RetroMAE (dense) \cite{xiao2022retromae}, and SPLADEv3 (learned sparse) \cite{lassance2024spladev}. The initial retrievals are re-ranked using a DeBERTa-v3 \cite{lassance2022naver} cross-encoder. The analysis demonstrates that as retrieval quality improves, LLM performance increases significantly across various models. Notably, re-ranking with SPLADEv3 and DeBERTa-v3 consistently achieves the best results across datasets and metrics. This underscores the critical role that high-quality retrieval plays in determining overall RAG system effectiveness, suggesting that IR-focused LLMs could be a valuable asset in enhancing generation performance.

\paragraph{Impact of the Generator} The BERGEN study \cite{rau2024bergen} compares the performance of LLMs with gold passages (Oracle) against closed-book settings without retrieval, as shown in Figure \ref{fig:regencomp02}. Surprisingly, the experiments do not reveal a straightforward relationship between model size and the performance gains from retrieval. For instance, smaller models like LLaMA2-7B benefit more from retrieval than larger models like LLaMA2-70B. In fact, LLaMA2-7B with retrieval outperforms LLaMA2-70B in a closed-book setting, suggesting that retrieval augmentation can make smaller models more competitive. Similarly, results from the eRAG experiments in Figure \ref{fig:regencomp03} indicate that varying LLM sizes (e.g., T5-small vs. T5-base) does not significantly affect the correlation between eRAG and downstream performance. These findings highlight that retrieval quality has a more substantial impact on RAG performance than the choice of generator, reinforcing the notion that investing in better retrieval strategies often yields more benefits than relying solely on larger LLMs.

\section{Challenges and Future Directions}
\label{sec:future}
The evolving landscape of RAG systems faces significant challenges that impact the quality of generated outputs, system efficiency, and the integration of multimodal data. As these systems become more prevalent across a range of applications, addressing these challenges is essential for improving their effectiveness and scalability. 

\subsection{Retrieval Quality}

The quality of retrieval is fundamental to any effective RAG system, directly influencing the relevance and accuracy of the generated content \cite{herrera‐viedma2006evaluating, zhong2012effective, wang2022webformer, salemi2024evaluating}. Current retrieval methods, however, frequently struggle with issues like noise, irrelevant documents, and fragmented information, all of which compromise the generation process.

\paragraph{Noise Robustness}
Irrelevant or misleading documents within the retrieved set can introduce noise, leading to hallucinations or unreliable answers. This challenge highlights the need for more sophisticated filtering and context-aware retrieval methods that can better differentiate relevant from irrelevant content. However, Cuconasu et al. \cite{cuconasu2024power} present an interesting perspective by showing that, under certain conditions, the inclusion of irrelevant documents can enhance overall accuracy. This finding challenges conventional retrieval strategies and suggests the potential for developing specialized approaches that strategically integrate noise within the retrieval process.

\paragraph{Negative Rejection}
When retrieval fails to return relevant results, models often attempt to generate responses regardless, increasing the risk of incorrect outputs. This issue is particularly problematic when queries are poorly expressed or lack sufficient context, making it difficult for retrieval models to surface relevant documents. Techniques like generating a pseudo-document that captures the query's essence, as demonstrated by HyDE \cite{gao2023precise}, can help bridge this gap. By allowing retrieval systems to find more relevant documents even from suboptimal queries, HyDE improves retrieval accuracy, albeit with a trade-off in computational cost. Future research could focus on optimizing this process to balance improved retrieval accuracy with reduced latency.

\paragraph{Information Integration}
Complex queries often require synthesizing information from multiple documents, yet fragmented or conflicting information can result in incoherent or incomplete answers. Pre- and post-retrieval techniques play a critical role in addressing this challenge. Enhancing retrieval granularity and incorporating techniques like entity-level retrieval and re-ranking can improve the cohesiveness of retrieved documents. However, many post-retrieval methods, as investigated by Zhu et al. \cite{zhu2023large}, rely heavily on calling LLM APIs, which incurs significant costs. Exploring alternatives such as knowledge distillation to lightweight models could offer more scalable solutions, making advanced retrieval strategies more practical in online settings. \\

Recent research highlights the development of generative models for search as a promising direction for improving retrieval quality. Models like GERE \cite{chen2022gere} and PARADE \cite{li2024parade} enhance document re-ranking and fact verification by directly generating relevant document titles or evidence sentences. Fine-tuning pre-trained models like RankT5 \cite{zhuang2023rankt} for ranking-specific tasks has also demonstrated potential in boosting out-of-domain performance, which is crucial for generalizing RAG systems across diverse contexts.

\subsection{System Efficiency}

System efficiency remains a significant bottleneck, especially as RAG systems scale to handle large datasets and real-time applications. The multi-step nature of RAG workflows—including query classification, retrieval, re-ranking, and generation—adds complexity and latency, which can hinder overall performance.

\paragraph{Latency in Retrieval Processes}
As document collections grow, retrieval and re-ranking processes increasingly become sources of latency. Lightweight search methods and hybrid retrieval approaches that combine sparse and dense techniques offer potential solutions by balancing speed and accuracy. For example, indexing, a traditionally resource-intensive process, has seen innovations through differentiable search indices such as DSI \cite{tay2022transformer} and SEAL \cite{bevilacqua2022autoregressive}. These methods integrate retrieval within Transformer models, enabling direct mapping of text queries to document identifiers and thereby improving both performance and retrieval efficiency.

\paragraph{Computational Costs}
The introduction of deep learning-based re-ranking models like monoT5 \cite{nogueira2020document} and RankLLaMA \cite{DBLP:conf/sigir/MaWYWL24} brings significant computational overhead, particularly in scenarios requiring iterative reasoning. Future research could focus on optimizing these models or developing retrieval pruning techniques that reduce the number of documents passed to the generation phase without sacrificing performance \cite{xiong2023xrr}.

\paragraph{Modular Workflow Optimization}
The complexity of RAG systems often stems from interdependencies between components like chunking strategies, embedding models, and re-ranking algorithms. Modular designs that enable independent optimization of each step while accounting for cross-component interactions are key to enhancing system throughput \cite{gao2024modular}. Advanced chunking methods and hybrid search strategies could offer trade-offs that maximize both retrieval precision and speed. An example is the Hybrid with HyDE \cite{wang2024searching} approach, which integrates both sparse and dense retrieval to capture relevant documents from both lexical and semantic perspectives.

\subsection{Multimodal RAG}

The expansion of RAG systems to support multimodal data—encompassing text, images, and audio—presents new challenges. Integrating diverse modalities requires not only effective retrieval but also seamless alignment and generation across different data types.

\paragraph{Cross-Modal Alignment}
Aligning multimodal documents with text-based queries remains a core challenge. The complexity of mapping diverse data types into a unified retrieval framework necessitates improved cross-modal retrieval strategies capable of simultaneously handling text, image, and potentially video or audio data.

\paragraph{Coherent Multimodal Generation}
Generating responses that meaningfully integrate information from multiple modalities is another difficult task. Advanced generation models capable of reasoning across different modalities are required to produce outputs that are both contextually relevant and visually coherent. \\

Recent advancements in multimodal RAG, such as MuRAG \cite{chen2022murag}, REVEAL \cite{hu2023reveal}, and Re-ViLM \cite{yang2023revilm}, have shown potential in incorporating multimodal retrieval and generation into real-world applications like visual question answering \cite{chen2023reimagen}, image captioning \cite{sarto2022retrievalaugmented}, and text-to-audio generation \cite{yuan2024retrievalaugmented}. Moving forward, research will likely focus on refining these techniques, especially in scaling multimodal retrieval to handle larger datasets and more complex queries. Extending retrieval capabilities to include more diverse media types, such as video and speech, also represents a promising direction for the continued evolution of RAG systems.

\section{Conclusions}
\label{sec:conclusions}
In this paper, we have presented a comprehensive framework for understanding the RAG domain, highlighting its significance in enhancing the capabilities of  LLMs. Through a structured overview of RAG, categorizing various methods, and an in-depth analysis of its core technologies and evaluation methods, this study illuminates the path for future research. It identifies crucial areas for improvement and outlines potential directions for advancing RAG applications, especially in textual contexts. This survey aims to elucidate the core concepts of the RAG field from a retrieval perspective, and it is intended to facilitate further exploration and innovation in the accurate retrieval and generation of information.

\begin{acks}
This research is supported by the Natural Sciences and Engineering Research Council (NSERC) of Canada.
\end{acks}

\bibliographystyle{ACM-Reference-Format}
\bibliography{main}

\end{document}